\documentclass[11pt,a4paper]{article}
\pdfoutput=1

\usepackage[utf8]{inputenc}
\usepackage[english]{babel}
\usepackage{chicago}

\usepackage[round]{natbib}
\usepackage[margin=1in]{geometry}

\usepackage{graphicx, subfig, caption, a4wide, booktabs, fancyvrb, upquote, longtable, multirow, enumitem, ifsym, hyperref}
\usepackage{amsmath, amsfonts, amssymb}
\usepackage{bm}
\usepackage{url}

\usepackage{afterpage}

\usepackage{color}
\definecolor{darkgrey}{rgb}{0.2,0.2,0.2}
\usepackage{fancyvrb}
\DefineVerbatimEnvironment{Code}{Verbatim}
  {fontsize=\normalsize,
   formatcom=\color{darkgrey},
   baselinestretch=1}
   
\renewcommand{\hat}[1]{\widehat{#1}}

\makeatletter
\newcommand\code{\bgroup\@makeother\_\@makeother\~\@makeother\$\@codex}
\def\@codex#1{{\normalfont\ttfamily\hyphenchar\font=-1 #1}\egroup}
\makeatother

\usepackage{fancyvrb}
\DefineVerbatimEnvironment{Code}{Verbatim}
  {fontsize=\footnotesize,
   formatcom=\color{darkgrey},
   baselinestretch=0.9}
   
\begin{document}


\title{Model-Based Clustering with Data Correction for Removing
  Artifacts in Gene Expression Data}
\author{%
  William Chad Young \\ University of Washington \and
  Ka Yee Yeung \\ University of Washington Tacoma \and
  Adrian E. Raftery \\ University of Washington
  \thanks{William Chad Young is a PhD student,
    Department of Statistics, University of Washington, Box 354322,
    Seattle, WA 98195-4322, USA; Email: wmchad@uw.edu. Ka Yee Yeung is
    Associate Professor, Institute of Technology, University of
    Washington - Tacoma, Campus Box 358426 1900 Commerce Street
    Tacoma, WA 98402, USA; Email: kayee@uw.edu. Adrian E. Raftery is
    Professor of Statistics and Sociology, Department of Statistics,
    University of Washington, Box 354322, Seattle, WA 98195-4322, USA;
    Email:  raftery@uw.edu. This research was supported by NIH grants
    U54-HL127624, R01-HD054511 and R01-HD070936. Computational
    resources provided by Microsoft Azure. The authors thank Ling-Hong
    Hung, Mario Medvedovic and Aravind Subramanium for helpful
    discussions.}}  \date{\today}
\maketitle

\begin{abstract}
  The NIH Library of Integrated Network-based Cellular Signatures
  (LINCS) contains gene expression data from over a million
  experiments, using Luminex Bead technology. Only 500 colors are used
  to measure the expression levels of the 1,000 landmark genes
  measured, and the data for the resulting pairs of genes are
  deconvolved. The raw data are sometimes inadequate for reliable
  deconvolution leading to artifacts in the final processed
  data. These include the expression levels of paired genes being
  flipped or given the same value, and clusters of values that are not
  at the true expression level. We propose a new method called
  model-based clustering with data correction (MCDC) that is able to
  identify and correct these three kinds of artifacts simultaneously.
  We show that MCDC improves the resulting gene expression data in
  terms of agreement with external baselines, as well as improving
  results from subsequent analysis. \\

\end{abstract}

\newpage
\baselineskip=18pt


\section{Introduction}

One of the major challenges in molecular biology is that of identifying
gene regulatory networks. This problem addresses the central dogma of
molecular biology as put forth by Francis Crick, most simply stated as
``DNA makes RNA makes protein''\citep{crick1970}. Much effort has
been put into discovering the mechanics behind this statement with the
knowledge that understanding the genetic interactions within cells
will lead to a much fuller understanding of how organisms develop at a
cellular level, as well as how diseases such as cancer affect cells and
how they can be treated.

Recent improvements in gene expression measurement technologies, such
as microarrays \citep{ball2002} and RNAseq \citep{wang2009}, have
greatly increased the amount of data available for inferring these
networks. This explosion in data has led to the development of a
number of methods for inferring the underlying regulatory
network. These include stochastic methods such as mutual information
\citep{basso2005,faith2007,margolin2006,meyer2007}, linear models
\citep{gustafsson2009, lo2012, menendez2010, young2014}, and Bayesian
networks \citep{kim2003,murphy1999,zou2005} as well as
deterministic methods such as differential equations
\citep{bansal2006,dhaeseleer1999}. Some of these
methods model the network holistically, while others identify the most
probable regulatory relationships supported by the data.

The performance of any method is limited by the
quality of the data being used. This is true for data from
gene expression experiments due to a number of factors, from
variability in environmental conditions to uncertainties inherent in
the measurement technologies themselves \citep{liu2010}. 
The data used for inference have usually gone through a preprocessing
pipeline to adjust the data to be more amenable to analysis
\citep{sebastiani2003}. Examples of preprocessing steps include
log transformation of raw fluorescence values and quantile
normalization. Although these techniques are often helpful, they can
sometimes introduce artifacts into the data \citep{lehmann2013}. It is
important to identify these additional sources of variation and
correct them if possible, or if not to account for them in the assessment
of variability and uncertainty.

Our work is motivated by a gene expression dataset from the NIH
Library of Integrated Network-based Cellular Signatures (LINCS)
program. In this dataset, genes are paired in the experimental setup
and this leads to multiple issues in the processed data, including
clustering, switched expression values of the two genes, and assignment of
the same expression value to the two genes. We develop
a new method to fix these issues. The method is an extension of
model-based clustering that explicitly incorporates the expression
level swaps while simultaneously addressing the other problems in the
data. We call it model-based clustering with data correction, or
MCDC. We show that our method works well on simulated datasets. We
also show that it improves the gene expression data, both in terms of
agreement with an external baseline and in subsequent inference.

Section \ref{s2} describes the motivating data for our method. Section
\ref{s3} outlines our method, MCDC, as well as a practical EM
algorithm for implementation. In Section \ref{s4} we present a
simulation study, showing the our method is able to identify and
correct points which have been altered. Section \ref{s5} shows how
MCDC can be applied to our motivating data to improve the data overall
as well as improve subsequent analyses. Section \ref{s6} concludes
with a discussion.

\section{Data} \label{s2}

The Library of Integrated Network-based Cellular Signatures (LINCS)
program \citep{duan2014}, \url{http://lincsproject.org}, is funded by
the Big Data to Knowledge (BD2K) Initiative at the National Institutes
of Health (NIH) whose aim is to generate genetic and molecular
signatures off human cells in response to various perturbations. This
program includes gene expression, protein-protein interaction, and
cellular imaging data \citep{vempati2014}. These data allow
researchers to gain new insights into cellular
processes. \cite{vidovic2013} used the LINCS data to understand drug
action at the systems level. \cite{shao2013} studied kinase inhibitor
induced pathway signatures. Both \cite{chen2015} and \cite{liu2015}
looked at associating chemical compounds with gene expression
profiles.

The LINCS L1000 data is a vast library of gene expression profiles
that include over one million experiments covering more than seventy
human cell lines. These cell lines are populations of cells descended
from an original source cell and having the same genetic makeup, kept
alive by growing them in a culture separate from their original
source. The L1000 data include experiments using over 20,000 chemical
perturbagens, namely drugs added to the cell culture to induce changes
in the gene expression profile. In addition, there are genetic
perturbation experiments targeting a single gene to control its
expression level, either suppressing it (knockdown) or enhancing it
(overexpression). The LINCS L1000 data is publicly available for
download from \url{http://lincscloud.org} and from the Gene Expression
Omnibus (GEO) database with accession number GSE70138
\url{http://www.ncbi.nlm.nih.gov/geo/query/acc.cgi?acc=GSE70138}.

\subsection{Experimental design of the L1000 data}

Each individual L1000 experiment measures the expression levels of
approximately 1,000 landmark genes in the human genome. The goal of
the LINCS project is to capture the cells' response to
perturbations. Therefore, the project was designed to include a very 
large number of experiments, but this came at the expense of measuring 
only a limited number of selected landmark genes. 
These landmark genes were selected to cover as much of the
variation in cellular gene expression as possible. In each experiment,
the selected perturbation was applied and the cells were allowed to
culture for a specified period of time before the gene expression
levels were measured.

The L1000 experiments were performed using the Luminex Bead technology
\citep{peck2006, dunbar2006}, in which color-coded microspheres are
produced to attach to specific RNA sequences corresponding to a
landmark gene and fluoresce according to the amount of RNA produced as
that gene is expressed. Sets of beads for measuring the 1,000 landmark
genes were added to the solution for a single experiment along with
the perturbing agent. The gene expression levels were measured by
sampling the beads from solution and analyzing each bead using lasers
to both classify the bead, identifying the specific gene being
measured, and measure the fluorescence level, indicating the
expression level of the gene.

The L1000 experiments used only 500 bead colors to measure the
expression levels of the 1,000 landmark genes. This means that each
bead color had to do double duty, accounting for a pair of genes. These
gene pairs were selected to have different levels of
expression, and the beads for a pair were mixed in approximately a 2:1
ratio. This means that, ideally, when the beads are sampled a
histogram of fluorescence levels corresponding to gene expression is
created with two peaks, one of which has twice the number of
observations as the second peak.

\subsection{L1000 Data Preprocessing}

\begin{figure}[ht]
  \centering
  \includegraphics[width=.6\textwidth]{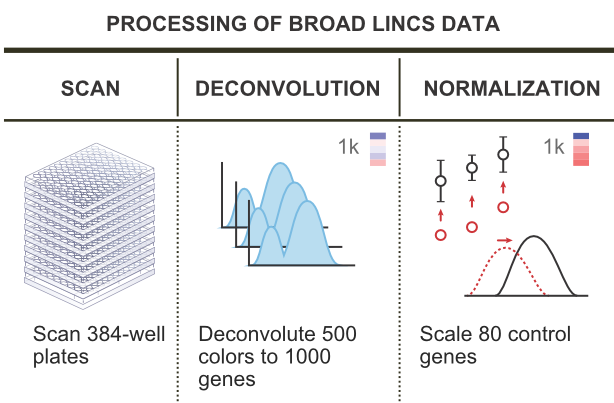}
\caption[L1000 data preprocessing pipeline] {
  The L1000 data preprocessing pipeline. The raw data are first measured
  from the beads in the experiments. Next, the data from each color of
  bead are deconvolved to assign expression values to the two genes
  which share that bead color. Finally, the data are normalized
  to yield directly comparable data across experiments. {\it Source:}
  Broad Institute LINCS cloud website (\url{http://lincscloud.org}).}
\label{fig:l1000proc}
\end{figure}

In order to facilitate statistical analysis on the L1000 data, the raw
bead fluorescence measurements are combined and transformed. First,
the measurements from many beads of the same color are deconvolved to
assign expression values to the appropriate pair of genes. The data
then goes through multiple normalization steps
\citep{liu2015,bolstad2003}. First, a set of genes are identified as
being stable across cell lines and perturbations, and these are used
to inform a power law transformation of all gene values. The
expression values are then quantile-normalized across sets of
experiments to make the distribution of expression levels the same for
all experiments. These steps are illustrated in Figure
\ref{fig:l1000proc}.

Although these data processing steps result in data that are more
amenable to statistical analysis, we have found that the deconvolution
step in particular introduces artifacts in the data. This can be seen
when we look at multiple experiments on the same cell line with the
same experimental conditions. If we look at a pair of genes that
share a bead color and form a scatterplot of their values across many
experiments, we see that these artifacts can take several forms.

\begin{figure}[ht]
  \centering
  \includegraphics[width=.48\textwidth]{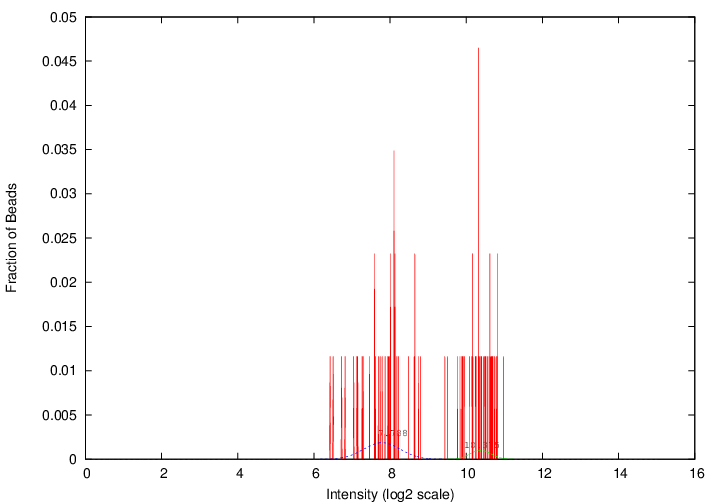}
  \includegraphics[width=.48\textwidth]{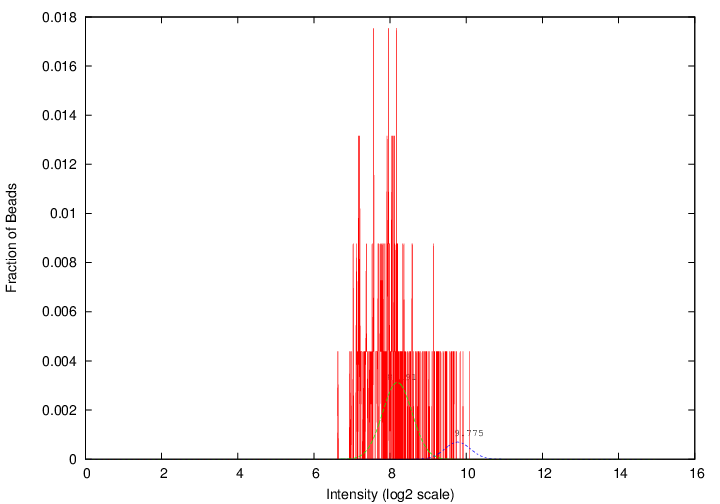}
  \caption[Bead fluorescence histograms]{Example histograms of raw
    bead fluorescence values for single bead colors. The plot on the
    left shows an example where the two peaks corresponding to the
    genes sharing the bead color are relatively easy to
    distinguish. The right plot shows an example where the
    deconvolution is difficult.}
  \label{fig:beadHist}
\end{figure}

First of all, two genes that are paired on the same bead color may
not be expressed at levels such that they are easily
distinguished. This can lead to assigning both genes the same value,
resulting in a clustering of data directly on the $x=y$
diagonal. Secondly, the deconvolution step, which uses a simple
$k$-means algorithm, can be misled if there are many beads sampled
with very low fluorescence values. This, combined with the quantile
normalization step, can lead to additional clusters that are not at
the true expression value. Finally, the deconvolution step can result
in assigning the expression levels of the genes incorrectly. That is,
the expression level of gene A of the pair on the same bead color is
sometimes assigned to gene B instead, and vice versa. Figure
\ref{fig:beadHist} shows examples of the raw bead data and illustrates
the difficulty of the deconvolution step.

\begin{figure}[ht]
  \centering
  \includegraphics[width=.6\textwidth]{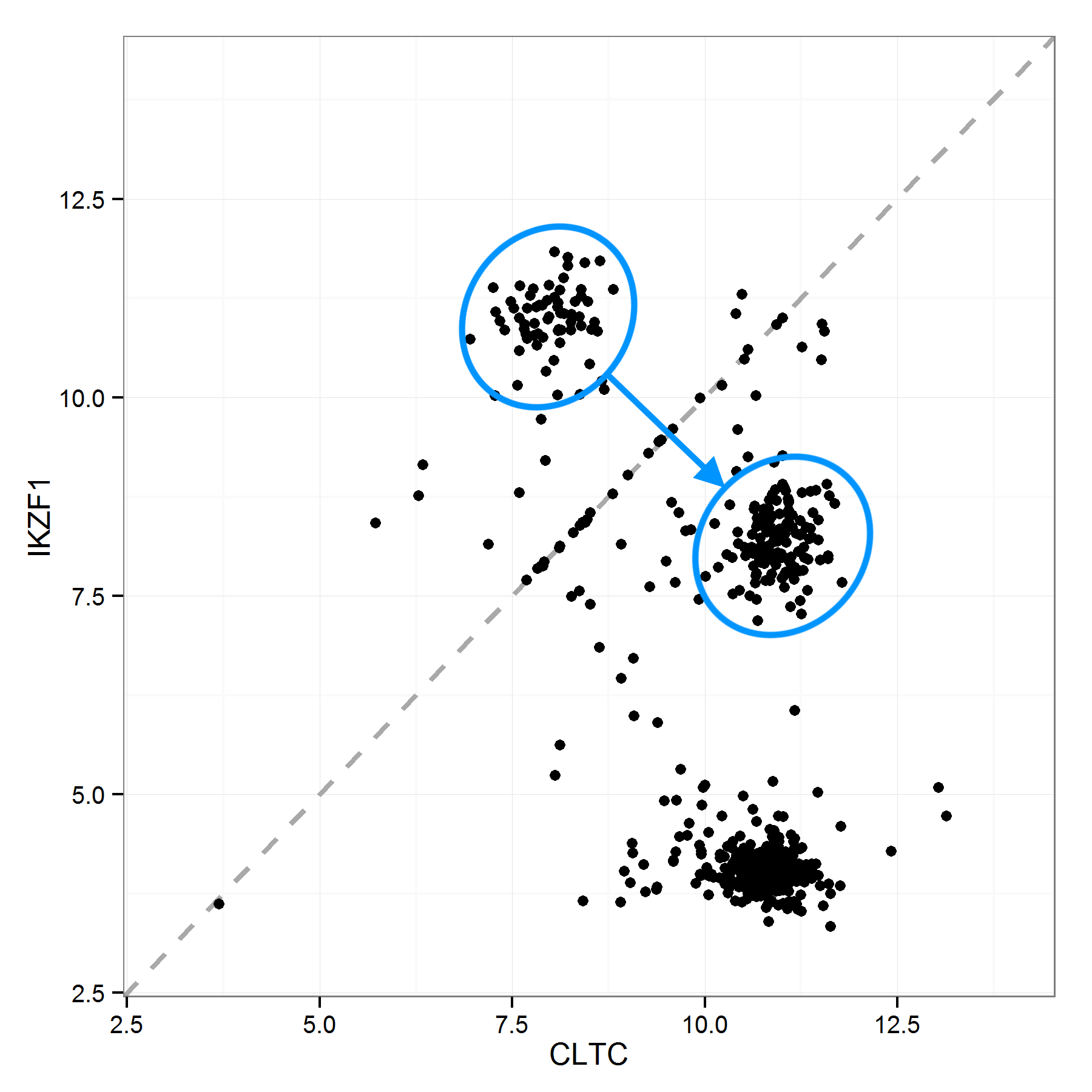}
  \caption[L1000 data showing examples of data artifacts.]
  {Expression levels on a log-base-2 scale for two paired genes, CLTC
    and IKZF1, measured on the same bead, in the L1000 data. Data
    artifacts include points directly on the diagonal, multiple
    clusters rather than a single one as may be expected, and flipping
    between the two circled clusters of points, with the CLTC value
    incorrectly assigned to IKZF1, and vice versa. The blue arrow
    shows the effect of data correction using MCDC.}
  \label{fig:flipEx}
\end{figure}

Figure \ref{fig:flipEx} shows examples of these three types of
artifact in the L1000 data. The figure shows the expression values for
two paired genes, CTLC and IKZF1. Each point shows the values
measured in a single experiment. All experiments in this dataset are
on the same cell line, A375, and are untreated, used as controls. As
such, we would expect a single cloud of observations centered around the 
point defined by the true expression values of the two genes. Instead, we see
several clusters of observations, as well as points lying on or very
near the diagonal. Note in particular the two circled sets of
points. These appear to be a single cluster in which some of the
points were flipped, with the expression values assigned to the wrong
genes. If we flip one set of points across the diagonal, it falls
directly on the other set. These phenomena occur with varying
frequency across all the gene pairs.

\section{Method} \label{s3}

We propose a method to detect and correct all three kinds of artifact
present in the LINCS L1000 data: the multiple clusters introduced by the
preprocessing pipeline, the erroneous assignment of the same expression 
value to paired genes, and flipping of the expression levels of paired genes.
This improves the quality of
the data and leads to better downstream analysis. We do this by
looking at the data holistically rather than reprocessing the raw
data, as in \citet{liu2015}. We adopt this holistic approach because
some of the artifacts are likely to persist even if the deconvolution
method is improved for individual gene pairs in isolation.

Our method is an extension of model-based clustering
\citep{Wolfe1970,banfield1993,MclachlanPeel2000,fraley2002}, which is
a model-based method for finding clusters by fitting a multivariate
Gaussian mixture model to the data. 
It has been found useful in many different contexts, including
geochemical analysis \citep{templ2008}, chemometrics
\citep{fraley2007}, and health studies \citep{flynt2015}. 
This method is well adapted to
estimating the expression levels because sometimes there are small
groups of points not in the main cloud around the true value, such as
the points on the diagonal in Figure \ref{fig:flipEx}.
Model-based clustering can identify these groups as clusters and
remove or downweight them, thus preventing them from contaminating the
estimation of the gene expression levels.

However, while model-based clustering is able to identify the clusters as well
as identifying outliers, it does not have a mechanism for identifying
particular points as flipped. Here we extend the model-based clustering
method to detect and take into account the flipping in the data. More
generally, it can be used for data with any invertible transformation
applied to a subset of the data. This extension allows us to use an
Expectation-Maximization (EM) algorithm commonly used to estimate
finite mixture models \citep{Dempster&1977,McLachlanKrishnan1997}.

\subsection{Model}

Suppose we have some multivariate data,
$\{\mathbf{x}_i: \; i = 1,\dots,N$, 
generated by a finite mixture of $G$ distributions $f_k$, $k =
1,\dots,G$ with probabilities $\tau_1,\dots,\tau_G$:
\begin{equation*}
  \label{eq:1}
  f(\mathbf{x}) = \sum_{k=1}^G \tau_k f_k(\mathbf{x} | \theta_k).
\end{equation*}
Suppose further that we do not observe $\mathbf{x}_i$, but rather
$\mathbf{y}_i$, a possibly transformed version of $\mathbf{x}_i$,
where the probability of a data point having been transformed can depend 
on the mixture component $k$ that $\mathbf{x}_i$ is drawn from:
\begin{equation*}
  \label{eq:2}
  \mathbf{y}_i = \left\{ \begin{array}{ll}
        \mathbf{x}_i & \text{with probability }\pi_k, \\
        \mathbf{T} \mathbf{x}_i & \text{with probability }(1-\pi_k).
      \end{array}\right.
\end{equation*}
Here, $\mathbf{T}$ is any invertible transformation that preserves the
domain of $\mathbf{x}$. Often, this may be represented as a matrix,
but it may also be a functional transformation (i.e. a component-wise
monotonic transformation). In the case of the L1000 data, this is just
the $2 \times 2$ matrix with zeros on the diagonals and ones on the
off-diagonals, switching the two values.

Given the transformation $\mathbf{T}$, the distribution of 
$\mathbf{y}_i$ can be written as follows:
\begin{equation*}
  \label{eq:3}
  f(\mathbf{y}_i | \bm{\tau}, \bm{\pi}, \bm{\theta}) = \sum_{k=1}^G
  \tau_k \left[ \pi_k f_k(\mathbf{y}_i | \theta_k) + (1-\pi_k) 
    f_k(\mathbf{T}^{-1}\mathbf{y}_i | \theta_k) \right].
\end{equation*}
To simplify the notation, define $f_{ik} \equiv
f_k(\mathbf{y}_i|\theta_k)$ and $f^-_{ik} \equiv
f_k(\mathbf{T}^{-1}\mathbf{y}_i|\theta_k)$. Then the distribution of 
$\mathbf{y}_i$ can be written
\begin{equation*}
  \label{eq:4}
  f(\mathbf{y} | \bm{\tau}, \bm{\pi}, \bm{\theta}) = \prod_{i=1}^n
  \sum_{k=1}^G \tau_k \left[ \pi_k f_{ik} + (1-\pi_k) f^-_{ik}
  \right]. 
\end{equation*}

\subsection{EM Algorithm}

We estimate this model by maximum likelihood using the EM algorithm.
We formulate this as a missing data problem where the complete data
are $\{\mathbf{y}_i, \mathbf{z}_i, \xi_i\}$. Here, $\mathbf{z}_i =
(z_{i1},\dots,z_{iG})$ and $\xi_i$ are unobserved labels, with
\begin{eqnarray*}
  z_{ik} = \left\{ \begin{array}{ll}
      1 & \text{if }\mathbf{y_i} \text{ belongs to group }k, \\
      0 & \text{otherwise},
    \end{array}\right.
\end{eqnarray*}
and
\begin{eqnarray*}
  \xi_i = \left\{ \begin{array}{ll}
      0 & \text{if }\mathbf{y_i} \text{ has been transformed}, \\
      1 & \text{otherwise}.
    \end{array}\right.
\end{eqnarray*}
Then the complete-data log-likelihood is
\begin{eqnarray*}
  \ell(\bm{\theta},\bm{\tau},\mathbf{z},
  \bm{\pi},\bm{\xi}|\mathbf{y}) = \sum_{i=1}^n \sum_{k=1}^G z_{ik}
  \left[ \right. \xi_i \log \left(\pi_k \tau_k f_{ik}
    \right) + \left. (1-\xi_i) \log \left((1-\pi_k) \tau_k
      f^-_{ik} \right) \right].
\end{eqnarray*}
We can also write down the joint distribution of $\mathbf{z}_i$ and
$\bm{\xi}_i$ given $\mathbf{y}_i$ and $\bm{\theta}$:
\begin{eqnarray}
  \label{eq:5}
  f(\mathbf{z}_i, \bm{\xi}_i | \mathbf{y}_i, \bm{\theta}) =
  \frac{1}{f(\mathbf{y}_i|\bm{\tau}, \bm{\pi}, \bm{\theta})}
  \prod_{k=1}^G \left[ (\tau_k \pi_k f_{ik})^{\xi_i} \cdot (\tau_k
    (1-\pi_k) f_{ik}^-)^{(1-\xi_i)}\right]^{z_{ik}}. 
\end{eqnarray}
For the E-step of the algorithm, we need to calculate
the expected complete-data log-likelihood, namely
\begin{eqnarray*}
  \label{eq:6}
  Q(\bm{\theta} | \bm{\theta}^*) &=&
  E[\ell(\bm{\theta},\bm{\tau},\mathbf{z},
  \bm{\pi},\bm{\xi}|\mathbf{y}) | \mathbf{y}, \bm{\tau}^*, \bm{\pi}^*,
  \bm{\theta}^*], \\ 
  &=& \sum_{i=1}^n \sum_{k=1}^G E[z_{ik} \xi_i | \mathbf{y},
  \bm{\tau}^*, \bm{\pi}^*, \bm{\theta}^*] \log \left(\pi_k \tau_k
    f_{ik} \right) + \\ 
  && E[z_{ik} (1-\xi_i) | \mathbf{y}, \bm{\tau}^*, \bm{\pi}^*,
  \bm{\theta}^*] \log \left((1-\pi_k) \tau_k f^-_{ik} \right).
\end{eqnarray*}
From Equation (\ref{eq:5}), we have
\begin{eqnarray*}
  \label{eq:7}
  E[z_{ik} \xi_i | \mathbf{y}, \bm{\tau}^*, \bm{\pi}^*, \bm{\theta}^*]
  &=& \frac{\tau_k^* \pi_k^* f_{ik}}{f(\mathbf{y}_i | \bm{\tau}^*,
    \bm{\pi}^*, \bm{\theta}^*)}, \\
  E[z_{ik} (1-\xi_i) | \mathbf{y}, \bm{\tau}^*, \bm{\pi}^*, \bm{\theta}^*]
  &=& \frac{\tau_k^* (1-\pi_k^*) f_{ik}^-}{f(\mathbf{y}_i |
    \bm{\tau}^*, \bm{\pi}^*, \bm{\theta}^*)}. \\
\end{eqnarray*}
We have $z_{ik}\xi_i + z_{ik}(1-\xi_i) = z_{ik}$ and $\sum_{k=1}^G
z_{ik} = 1$. This leads to the following updates of the estimates
of $z_{ik}$ and $\xi_i$, which make up the E-step:
\begin{eqnarray*}
  \label{eq:8}
  \hat{z}_{ik} &=& \frac{\tau_k^* [\pi_k^* f_{ik} + (1-\pi_k^*
    f_{ik}^-)]} {f(y_i | \bm{\tau}^*, \bm{\pi}^*, \bm{\theta}^*)}, \\
  \hat{\xi}_i &=& \frac{\sum_{k=1}^G \tau_k^* \pi_k^* f_{ik}} {f(y_i
    | \bm{\tau}^*, \bm{\pi}^*, \bm{\theta}^*)}. 
\end{eqnarray*}

The M-step is then as follows:
\begin{eqnarray*}
  \label{eq:9}
  \hat{\tau}_k &\leftarrow& \frac{n_k}{n}, \\
  \hat{\pi}_k &\leftarrow& \frac{\sum_{i=1}^n \hat{z}_{ik}\hat{\xi}_i}{n_k},\\
  \hat{\bm{\mu}}_k &\leftarrow& \frac{\sum_{i=1}^n \hat{z}_{ik} (\hat{\xi}_i
    \mathbf{y}_i + (1-\hat{\xi}_i) \mathbf{T}^{-1}\mathbf{y}_i)}{n_k},
  \\
  n_k &\equiv& \sum_{i=1}^n \hat{z}_{ik}.
\end{eqnarray*}

To get the variance of the clusters, we follow the steps from
\citet{celeux1995}, modifying the within-cluster scattering matrix $W$
and the scattering matrix $W_k$ of cluster $k$ to be
\begin{eqnarray*}
  \label{eq:10}
  W &=& \sum_{k=1}^G \sum_{i=1}^n \hat{z}_{ik} \left[ \hat{\xi}_i (\mathbf{y}_i -
    \hat{\bm{\mu}}_k) (\mathbf{y}_i - \hat{\bm{\mu}}_k)' + (1-\hat{\xi}_i)
    (\mathbf{T}^{-1} \mathbf{y}_i - \hat{\bm{\mu}}_k)
    (\mathbf{T}^{-1}\mathbf{y}_i - \hat{\bm{\mu}}_k)' \right], \\
  W_k &=& \sum_{i=1}^n \hat{z}_{ik} \left[ \hat{\xi}_i (\mathbf{y}_i -
    \hat{\bm{\mu}}_k) (\mathbf{y}_i - \hat{\bm{\mu}}_k)' + (1-\hat{\xi}_i)
    (\mathbf{T}^{-1} \mathbf{y}_i - \hat{\bm{\mu}}_k)
    (\mathbf{T}^{-1}\mathbf{y}_i - \hat{\bm{\mu}}_k)' \right].
\end{eqnarray*}
These can then be used to calculate $\Sigma_k$ under different
variance models.

We iterate the EM steps until convergence, and this leads to a local
optimum of the log-likelihood \citep{Wu1983}. 
Although this is not guaranteed to be the global optimum, 
choosing starting values intelligently or doing
multiple restarts have been shown to lead to good solutions
\citep{Fraley1998,Biernacki&2003}.

Our model assumes that each cluster variance is independent. We select
the best number of clusters by running MCDC with the number of
clusters ranging from 1 to some maximum number of clusters (9 in our
case) and then comparing the BIC values for the resulting estimated
models \citep{fraley2002}.

For our gene expression data, we estimate the expression levels of a pair
of genes as the mean of the largest cluster (the cluster with
the most points assigned to it) found using the chosen model. This allows
us to ignore the artifactual clusters that result from the preprocessing
of the data in estimating the expression levels of the genes.

\section{Simulation Study}  \label{s4}

We now describe a simulation study in which data with some of the key
characteristics of the LINCS L1000 data were simulated. We simulated 
datasets with no clustering (i.e. one cluster), but where some of the
observations were flipped. We also simulated datasets with clustering
(two clusters), where some of the observations were flipped.

Finally, we simulated a dataset where no observations were flipped,
but instead some observations were rotated and scaled. This is to show
that the method can be effective when some of the data are perturbed in
ways other than flipping.

\subsection{Simulation 1: One Cluster With Flipping}

\begin{figure}[ht]
  \centering
  \includegraphics[width=.48\textwidth]{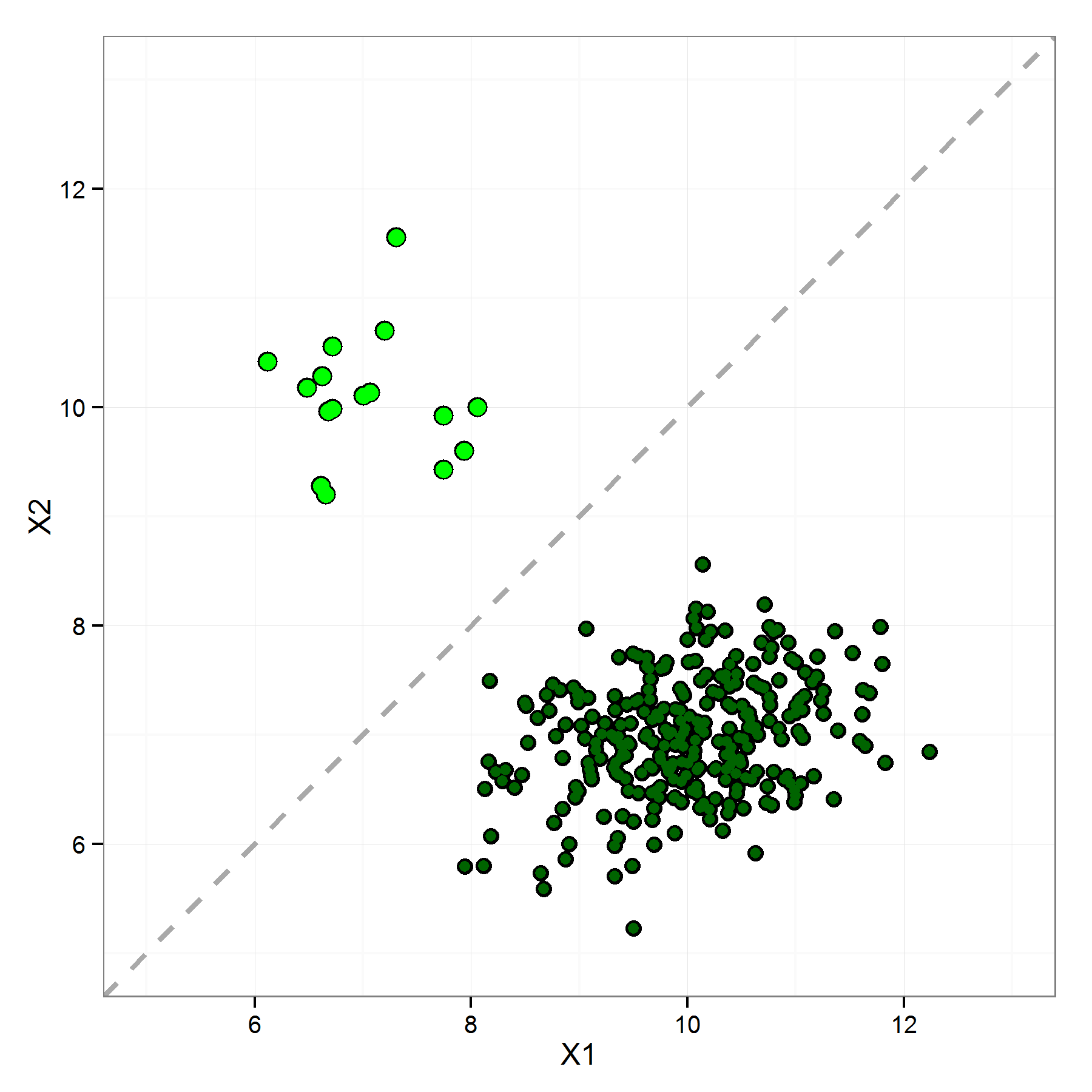}
  \includegraphics[width=.48\textwidth]{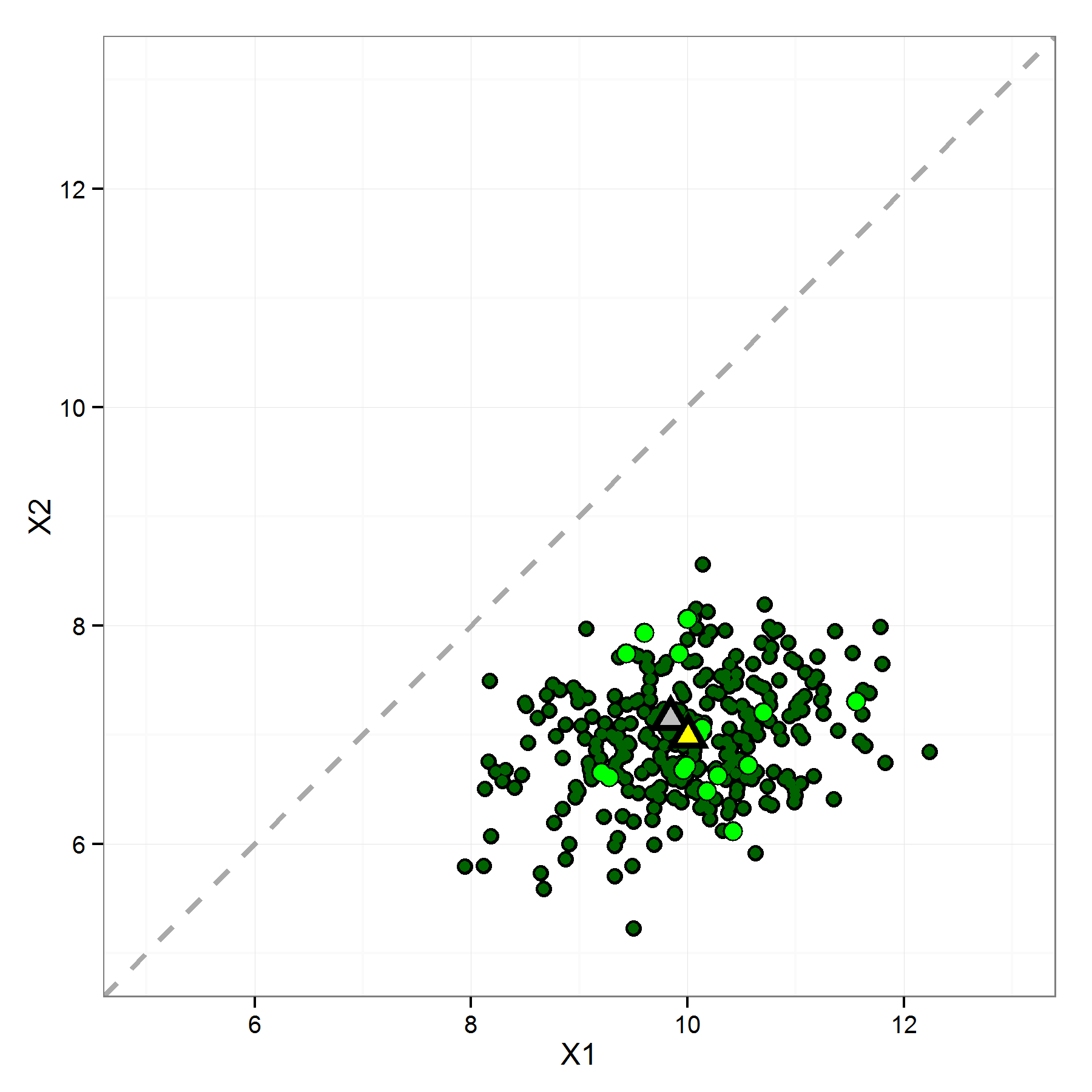}
  \caption[Simulation 1 - scatterplot and results from example
  dataset]{One Dataset from Simulation 1: One Cluster with Flipping.
    Left panel:  Original data with flipped data points.  Right panel:
    Data after correction by MCDC.  MCDC identified and corrected all
    the flipped points. The grey triangle is the mean of all the data,
    and the yellow triangle is the mean of the data after correction
    by MCDC.}
  \label{fig:sim1}
\end{figure}

Figure \ref{fig:sim1} is an example dataset from our first
simulation. This simulation represents what we see in the LINCS L1000
data in the best case, with no clustering or diagonal values 
(i.e. a single cluster), but with some flipping. 
For the simulation, we generated 100 datasets with 300
points each from the single cluster model with flipping probabilities
$(1-\pi)$ of 0.05 to 0.45 in increments of 0.05. We applied MCDC to each 
simulated dataset and counted the number of
times the correct number of clusters (one) was selected as well as the
percentage of the points correctly identified as flipped or
not. Finally, we looked at the inferred gene means compared to taking
the mean of the data without MCDC.

The correct number of clusters (i.e. one) was selected for all but one
of the 900 datasets, and in the one erroneous dataset out of 900
only a few points were in a second cluster. 
In 858 of the 900 datasets no errors were
made - all the points were correctly identified as being flipped or
not flipped. In three datasets, all with the highest probability of
flipping, namely 0.45, 
all the points were misidentified by being flipped to the
wrong side, while in one dataset (again with $\pi=0.55$), a single
large cluster with no flipping was identified. The remaining 38
datasets had one to three points out of 300 misidentified. 
All these misidentifications
make sense, since we expect rare cases where a point crosses the $x=y$
line as well as cases where more points are flipped when using a
flipping probability near 0.5.

\begin{figure}[ht]
  \centering
  \includegraphics[width=\textwidth]{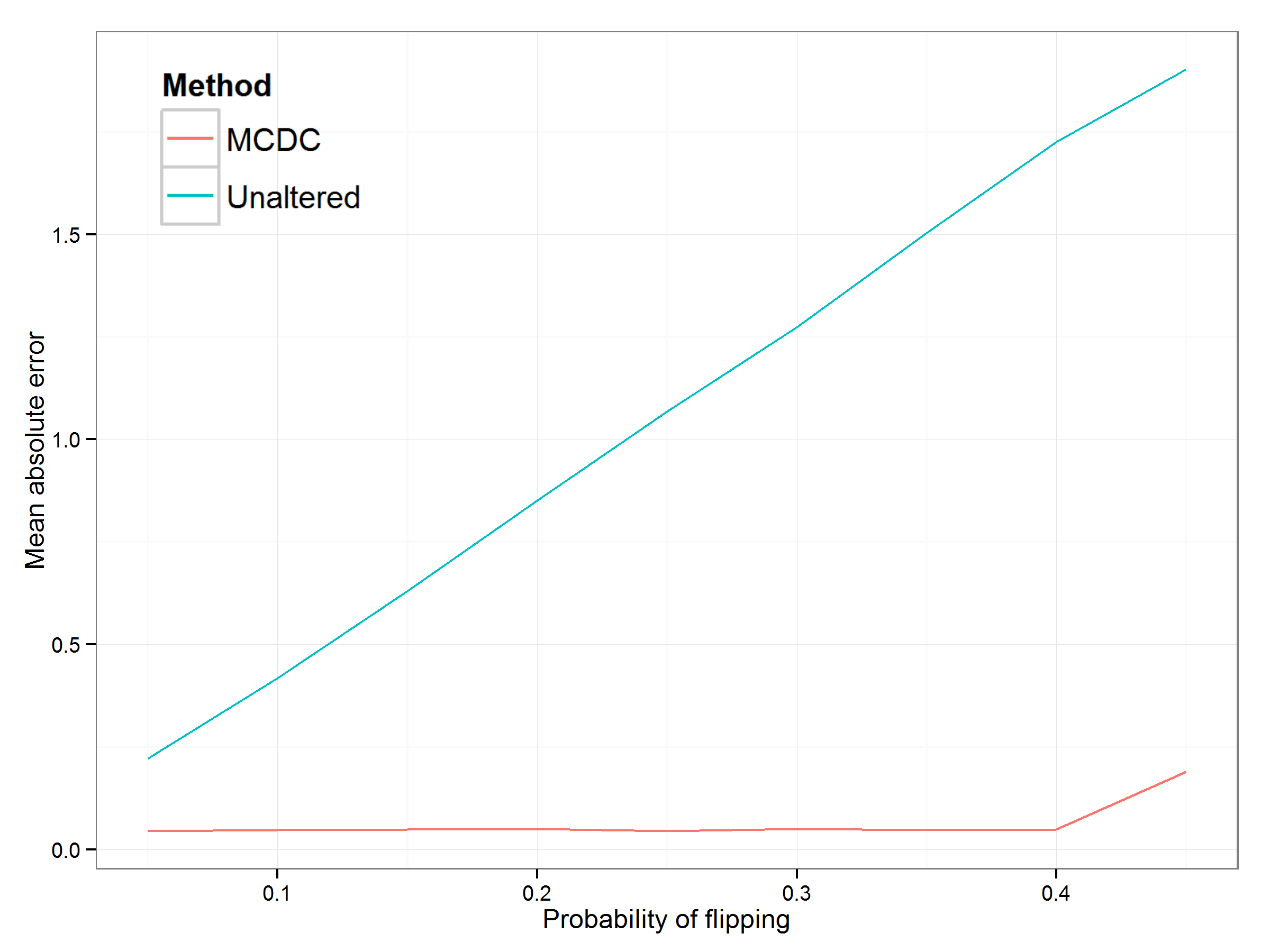}
  \caption[Simulation 1 - Mean Absolute Error comparison]{Simulation
    1: Mean Absolute Error in Inferred Mean.  The blue line is using
    unaltered data, while the red line is using the mean of the
    largest cluster found by MCDC.}
  \label{fig:sim1mean}
\end{figure}

Figure \ref{fig:sim1mean} and Table \ref{tab:sim1comp} show the
mean absolute error
in inferred mean using MCDC versus the unaltered
data. For each flipping probability, we calculated the mean absolute
error of the inferred mean from the true mean. MCDC did much better
than taking the unaltered mean in all cases, improving on the
unaltered data by a factor of 5 to 36, depending on the probability of
flipping.

\begin{table}[ht]
  \caption[Simulation 1 - Mean Absolute Error results]{Simulation 1:
    Mean Absolute Error (MAE) in Inferred Mean for Unaltered data and
    MCDC-corrected Data, as the probability of flipping increases. The
    MAE ratio is the ratio of mean absolute error using the unaltered
    data divided by the MAE using the MCDC-corrected data. Values
    greater than 1 indicate improvement by using MCDC.}
  \centering
  \begin{tabular}[ht]{c|ccc}
    \textbf{Probability of} & \textbf{Unaltered} &
    \textbf{MCDC} & \textbf{MAE} \\
    \textbf{Flipping} & \textbf{MAE} &
    \textbf{MAE} & \textbf{Ratio} \\
    \hline
    0.05 & 0.22 & 0.04 & 5 \\
    0.10 & 0.42 & 0.05 & 9 \\
    0.15 & 0.63 & 0.05 & 13 \\
    0.20 & 0.85 & 0.05 & 17 \\
    0.25 & 1.07 & 0.05 & 24 \\
    0.30 & 1.27 & 0.05 & 25 \\
    0.35 & 1.50 & 0.05 & 32 \\
    0.40 & 1.73 & 0.05 & 36 \\
    0.45 & 1.90 & 0.19 & 10 \\ \hline
  \end{tabular}
  \label{tab:sim1comp}
\end{table}

\subsection{Simulation 2: Two Clusters with Flipping}

\begin{figure}[ht]
  \centering
  \includegraphics[width=.48\textwidth]{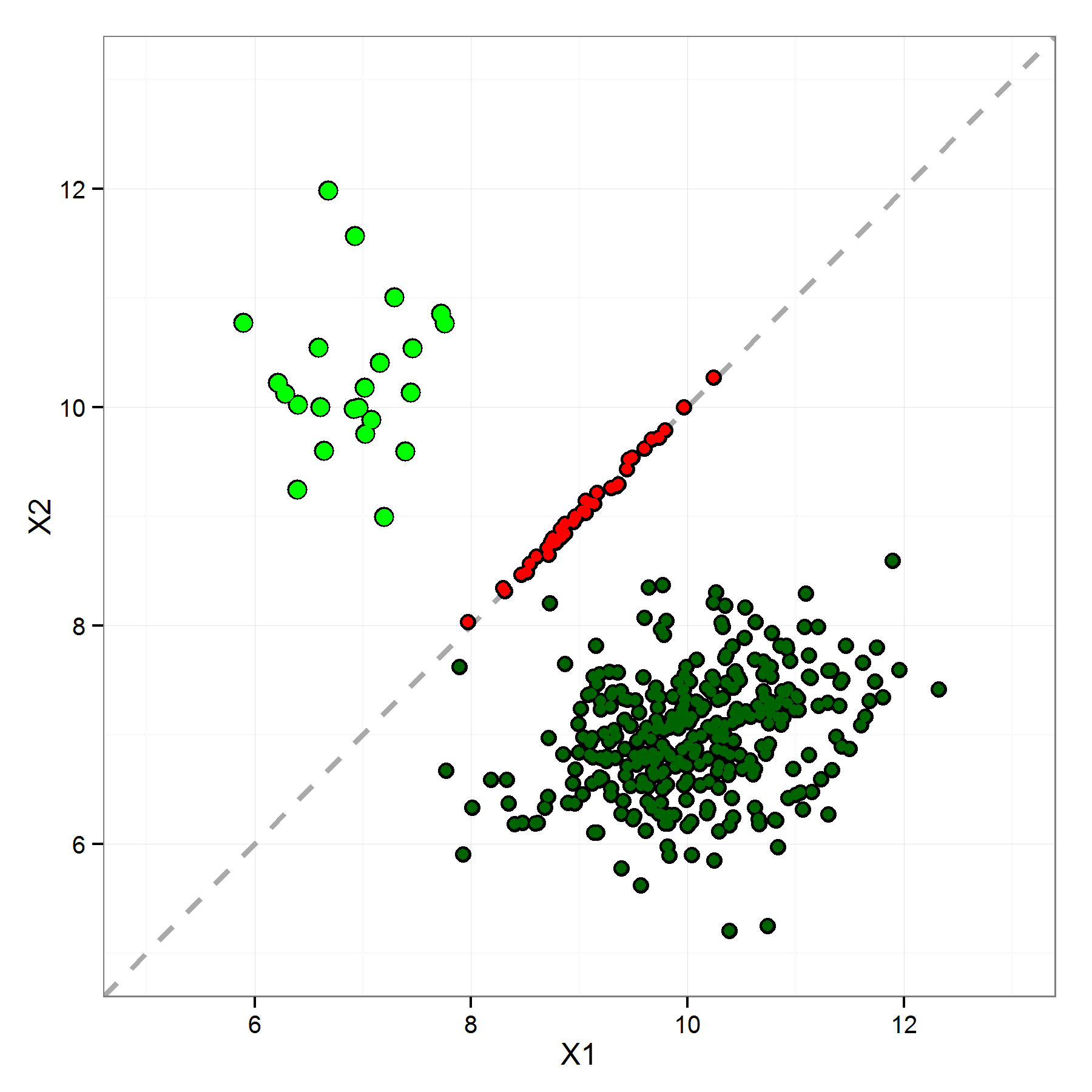}
  \includegraphics[width=.48\textwidth]{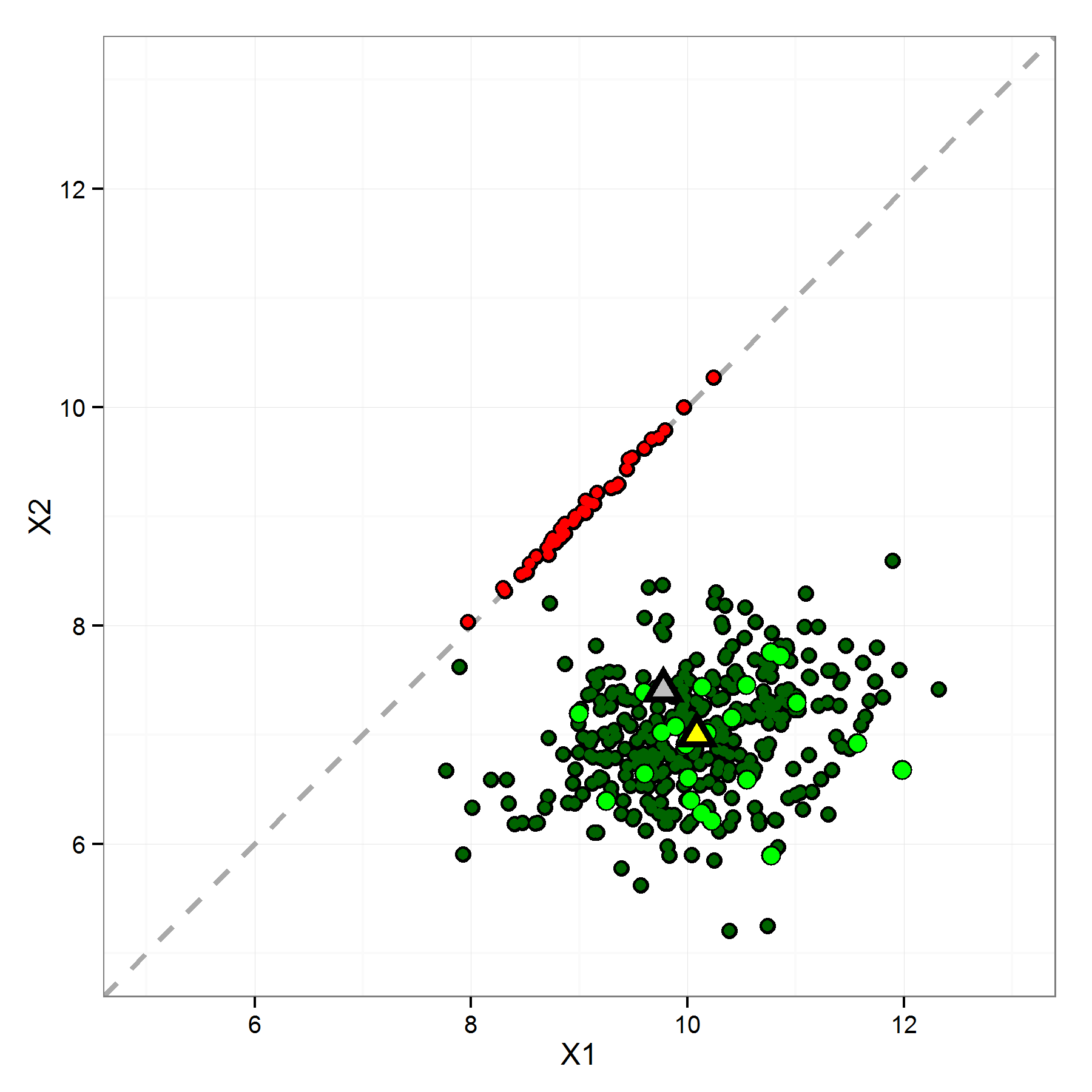}
  \caption[Simulation 2 - scatterplot and results from example
  dataset]{Simulated dataset 2: 2 clusters after flipping. MCDC
    correctly identifies the two clusters and the flipped points, as
    seen in the plot on the left. The grey triangle is the mean of all
    the data, which is moved from its true position due to the second
    cluster. The yellow triangle is the mean of the largest cluster
    found by MCDC and is much closer to the true value.}
  \label{fig:sim2}
\end{figure}

For the second simulation, we added a second cluster on the diagonal,
as demonstrated in Figure \ref{fig:sim2}. This reflects a common issue
we see in the L1000 data. When the data processing pipeline has
trouble differentiating between the two gene expression levels, it can
end up assigning them both the same value. For these data, we wanted
to see how well MCDC identified the ``good'' points (not the diagonal
cluster). Again, we used the mean of the largest cluster as the
inferred mean. For the simulation, we generated 100 datasets with 400
points each for the two-cluster model with flipping probabilities
$(1-\pi)$ of 0.05 to 0.45, and probability $\tau$ of a point being in
the true cluster 0.55 to 0.95, in increments of 0.05.

\begin{figure}[ht]
  \centering
  \includegraphics[width=\textwidth]{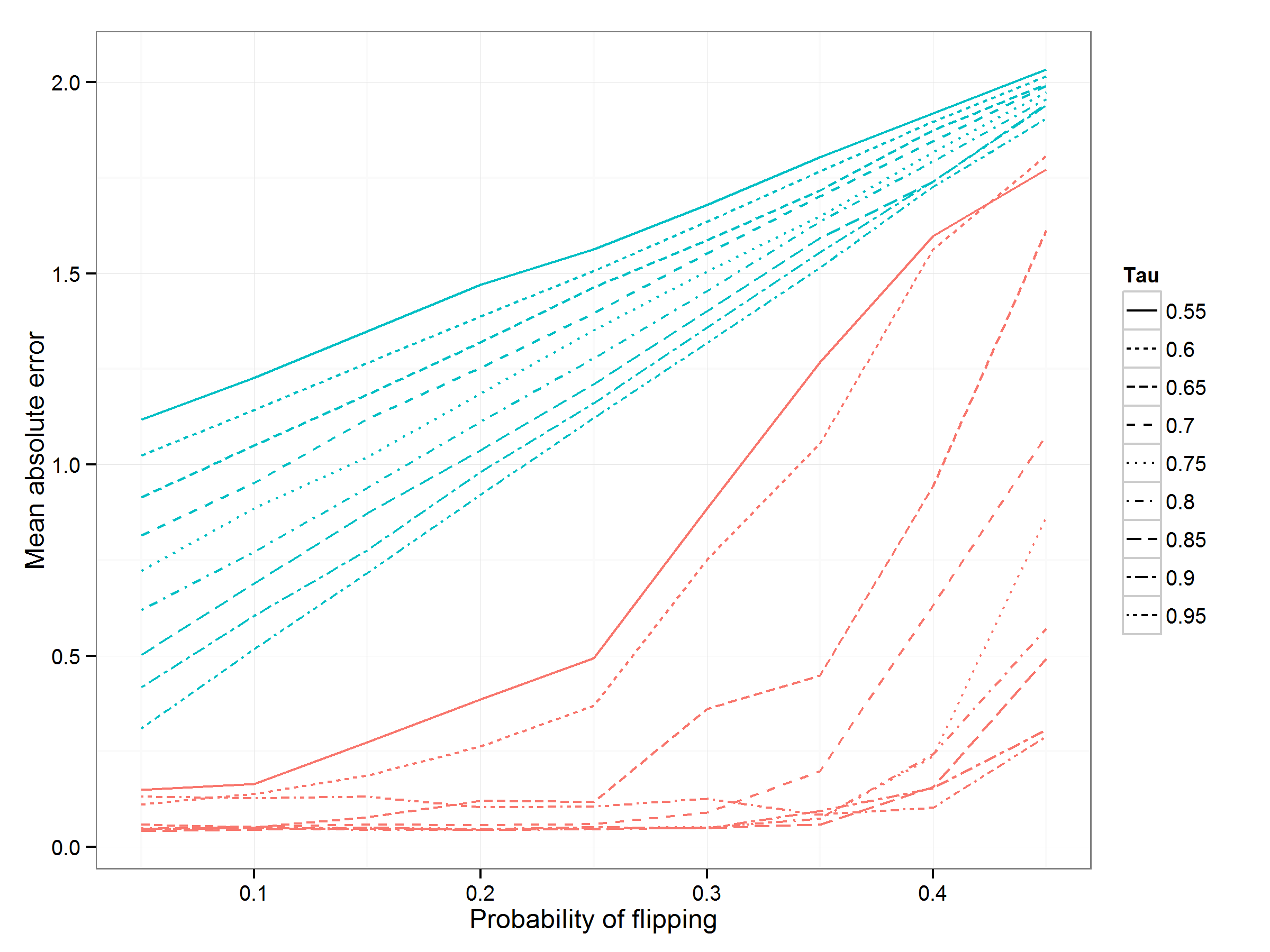}
  \caption[Simulation 2 - Mean Absolute Error comparison]{Mean
    absolute error in inferred mean comparison for simulation 2 when
    varying $\tau$, the probability that a point comes from the
    primary cluster.  The blue line is using unaltered data, while the
    red line is using the mean of the largest cluster found by MCDC.}
  \label{fig:sim2mean}
\end{figure}

\begin{figure}[ht]
  \centering
  \includegraphics[width=\textwidth]{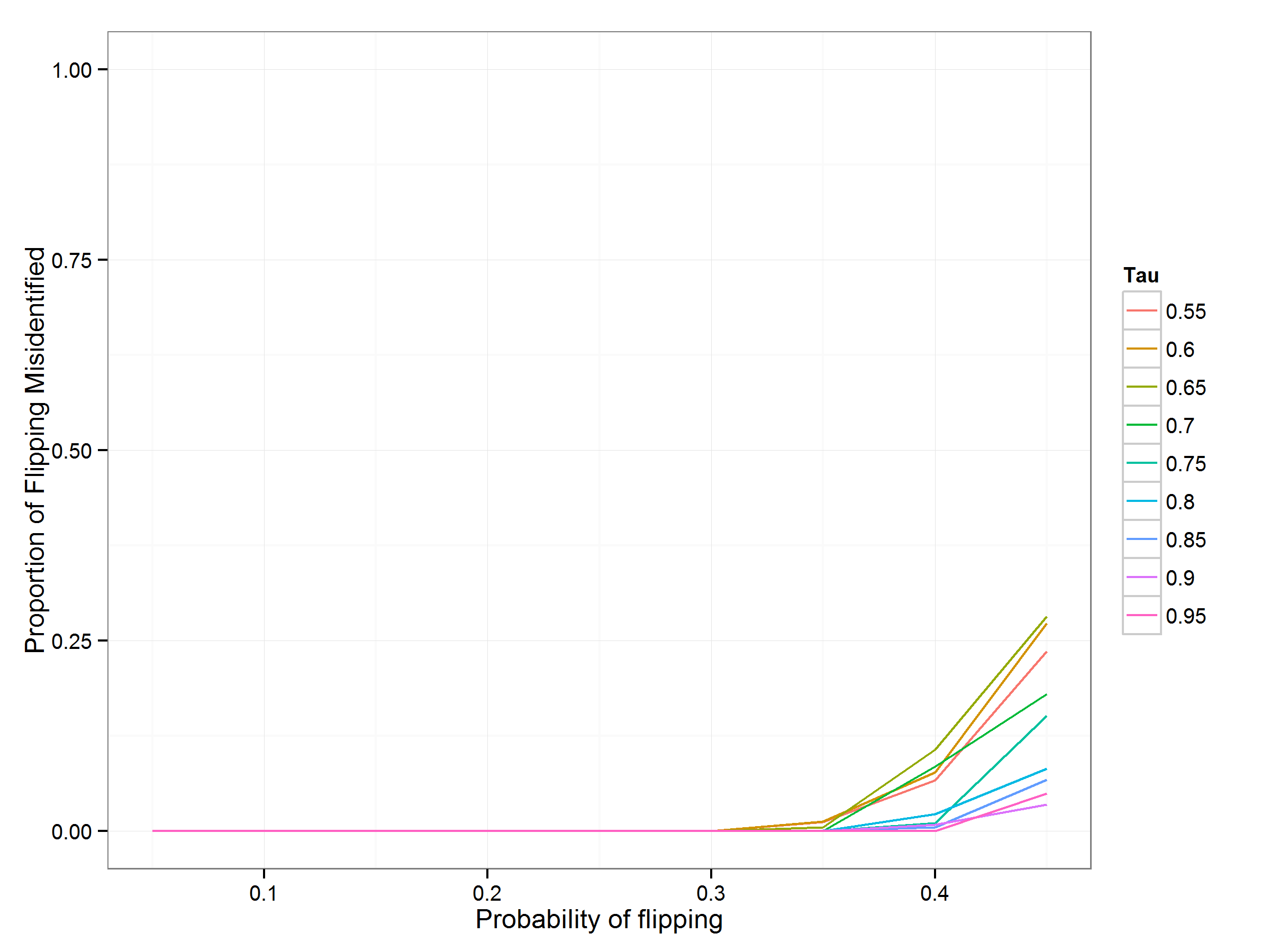}
  \caption[Simulation 2 - identification of flipped points]{Proportion
    of points correctly identified as flipped or not flipped for
    simulation 2 when varying $\tau$, the probability that a point
    comes from the primary cluster. When there is a high probability
    of flipping (near 0.5), there may be more points in the flipped
    cluster, leading to MCDC identifying it as the main cluster and
    thus misidentifying all points for a particular dataset.}
  \label{fig:sim2flip}
\end{figure}

\begin{figure}[ht]
  \centering
  \includegraphics[width=\textwidth]{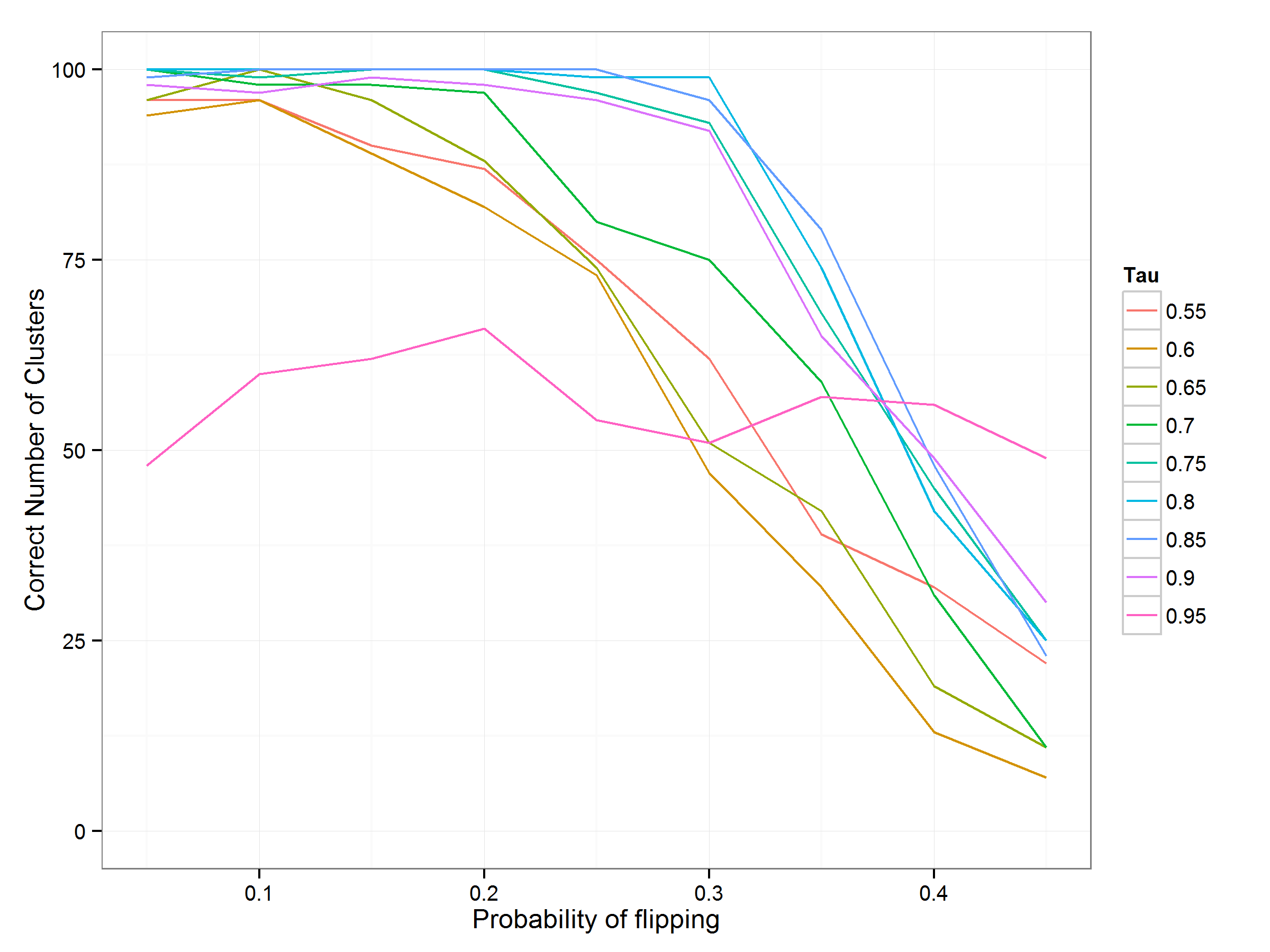}
  \caption[Simulation 2 - identification of correct number of clusters]{Number of
    datasets (out of 100) which identify 2 clusters as the best result
    for simulation 2 when varying $\tau$, the probability that a point
    comes from the primary cluster. When $\tau$ is high, MCDC may not
    identify the cluster on the diagonal properly due to the low
    number of points in that cluster.}
  \label{fig:sim2g}
\end{figure}

Figure \ref{fig:sim2mean} shows the comparison of mean absolute error
in the inferred mean. The MCDC results are better across the board,
although we see that as $\tau$ decreases, it is more likely to find
the diagonal cluster as the largest one. Figure \ref{fig:sim2flip}
shows that MCDC does well in identifying which points are flipped or
not. In Figure \ref{fig:sim2g}, we see that the correct number of
clusters is not generally identified as well as we might like. This
may be due to poor initialization of the algorithm and may be
corrected with multiple random initializations.

\subsection{Simulation 3: Three Clusters With Rotation and Scaling}
\begin{figure}[ht]
  \centering
  \includegraphics[width=.48\textwidth]{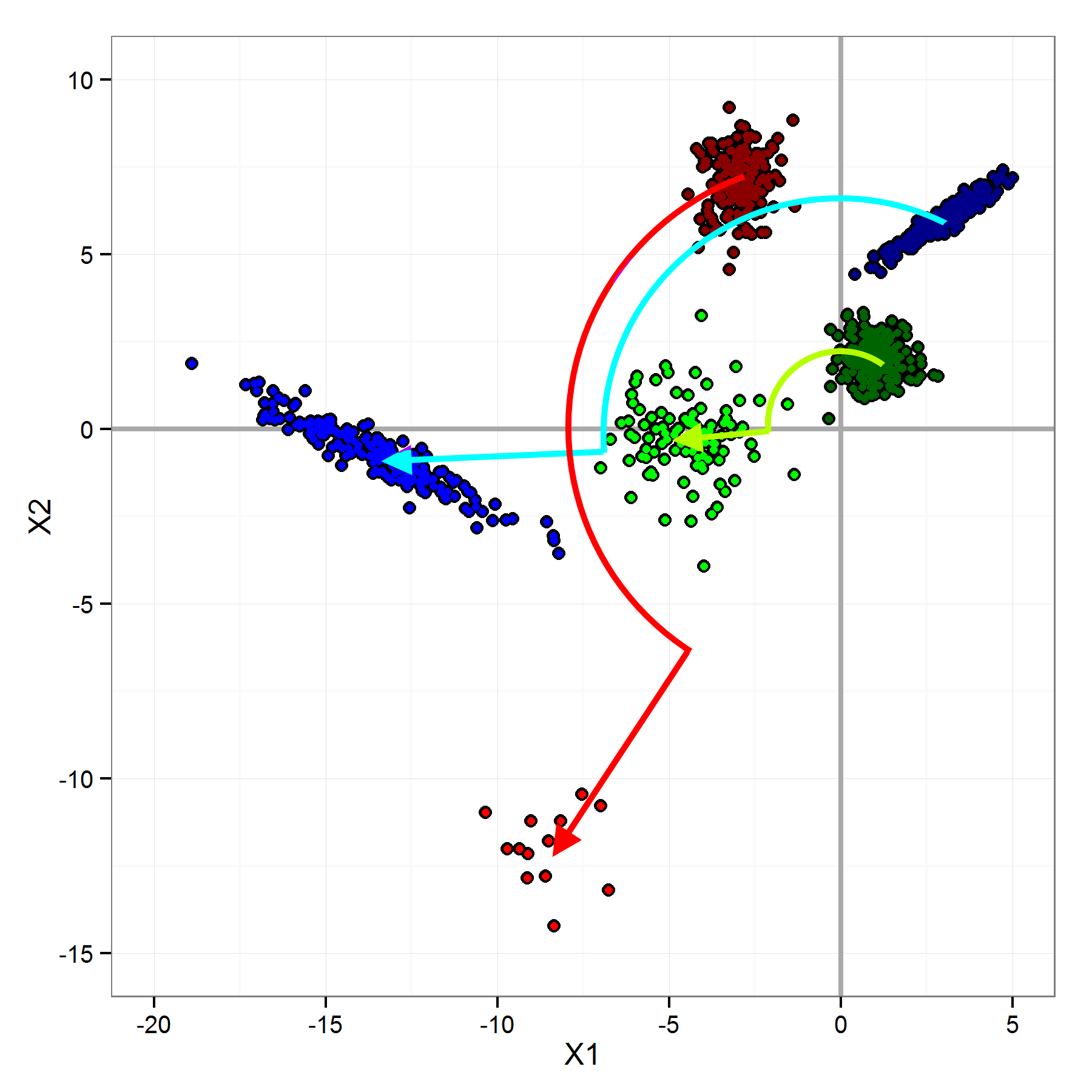}
  \includegraphics[width=.48\textwidth]{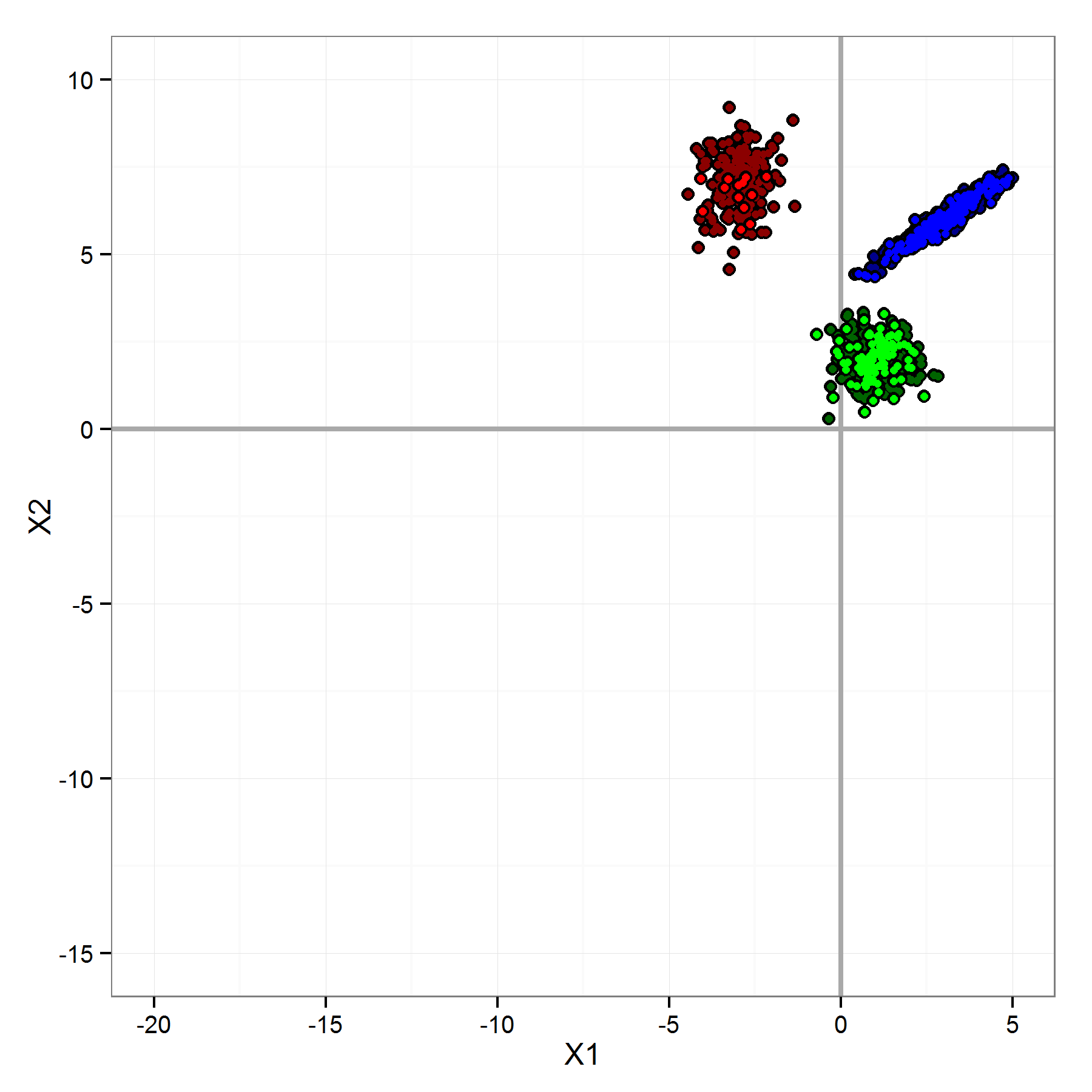}
  \caption[Simulation 3 - scatterplot and results from example
  dataset]{Simulation dataset 3: 3 clusters with rotation and
    scaling. A data point is transformed by rotating it $120^{\circ}$
    counter-clockwise around the origin and then scaling out from the
    origin by a factor of 2, as seen in the plot on the right. MCDC is
    able to identify the correct clusters and assign the transformed
    points back into the appropriate clusters, as on the left.}
  \label{fig:sim3}
\end{figure}

To show that MCDC can be applied to other
kinds of data errors than flipping, we also generated a dataset with
three clusters where the error process rotates and scales the
data points affected. 
In this more complex situation, we used $n=1000$ points split
among the three clusters with varying probabilities of
transformation. MCDC selected the correct number of clusters 
and correctly classified all the points. Figure \ref{fig:sim3} shows
the original and MCDC-corrected data. One caveat is that here,
as in the flipping situation, the data error process was known to
the MCDC algorithm.
  
\section{Application to LINCS L1000 Data}  \label{s5}
We applied MCDC to a portion of the LINCS L1000, namely the data from 
cell line A375, a human skin malignant melanoma cell line. 
We chose this cell line because it has good coverage in the L1000 data in
terms of the number of different perturbations applied. We looked at
improvement of the data in aggregate as well as improvement in a
specific inferential setting.

\begin{figure}[ht]
  \centering
  \includegraphics[width=.48\textwidth]{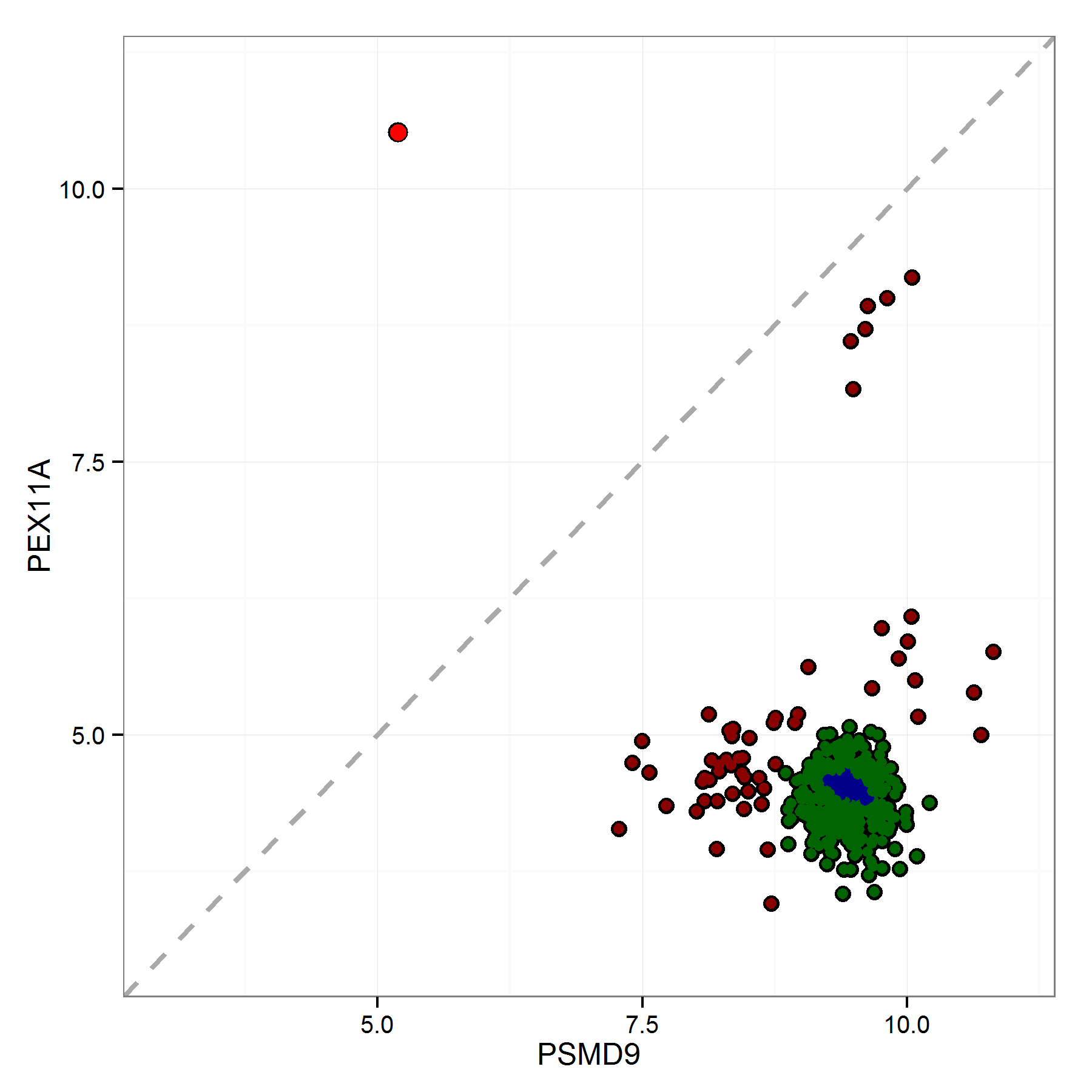}
  \includegraphics[width=.48\textwidth]{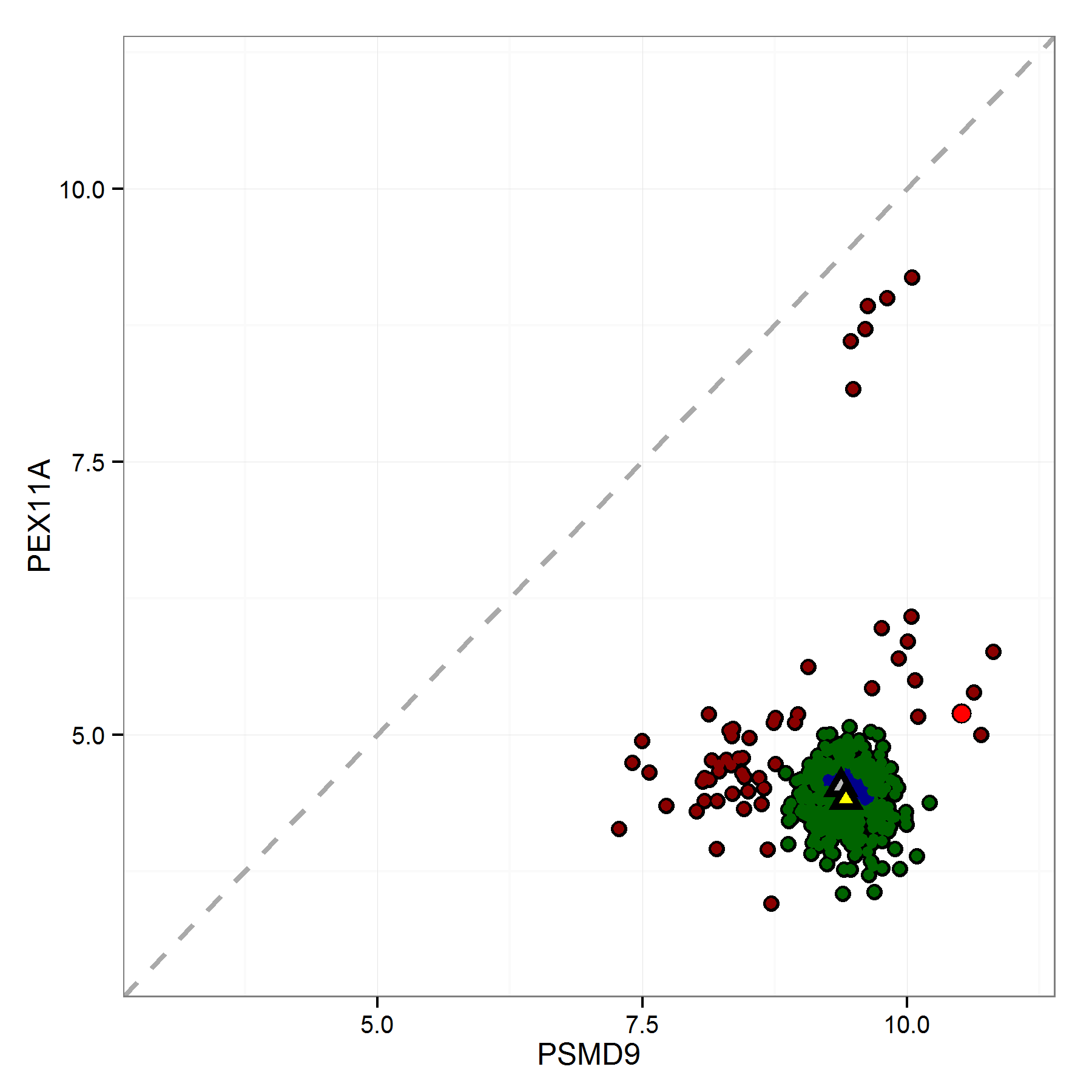}
  \caption[Results of MCDC on LINCS L1000 data - example 1]{Example 1
    showing results of applying MCDC to L1000 data. On the left is the
    data before correction, and on the right is that data after
    correction. Triangles indicate inferred mean - gray is the mean of
    all the data while yellow is the mean of the largest cluster found
    by MCDC.}
  \label{fig:example1}
\end{figure}

Figure \ref{fig:example1} shows the results of applying MCDC to one
gene pair in untreated experiments on cell line A375. Each point
corresponds to a single experiment, and most of the points fall in the
same region. However, there are a few points which appear to be
mirrored across the $x=y$ line, and we suspect that these points have
had the expression levels of the two genes switched. MCDC corrects
these points and we see that they do indeed fall within the main body
of points. 

Note that here the best solution involves three
components. This means that the distribution of the points may not be
strictly normal, but the components overlap such that they form one
contiguous cluster. 

\begin{figure}[ht]
  \centering
  \includegraphics[width=.48\textwidth]{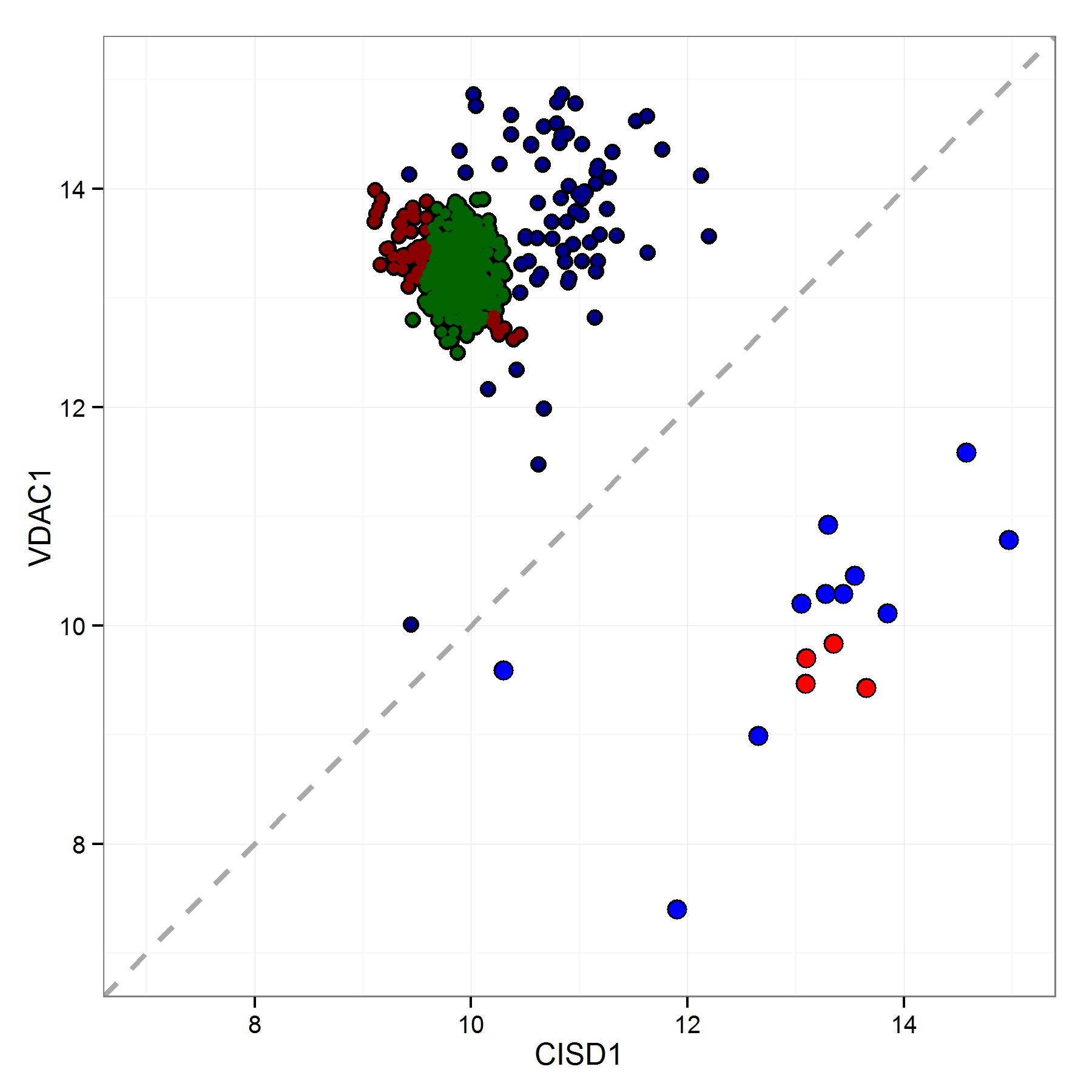}
  \includegraphics[width=.48\textwidth]{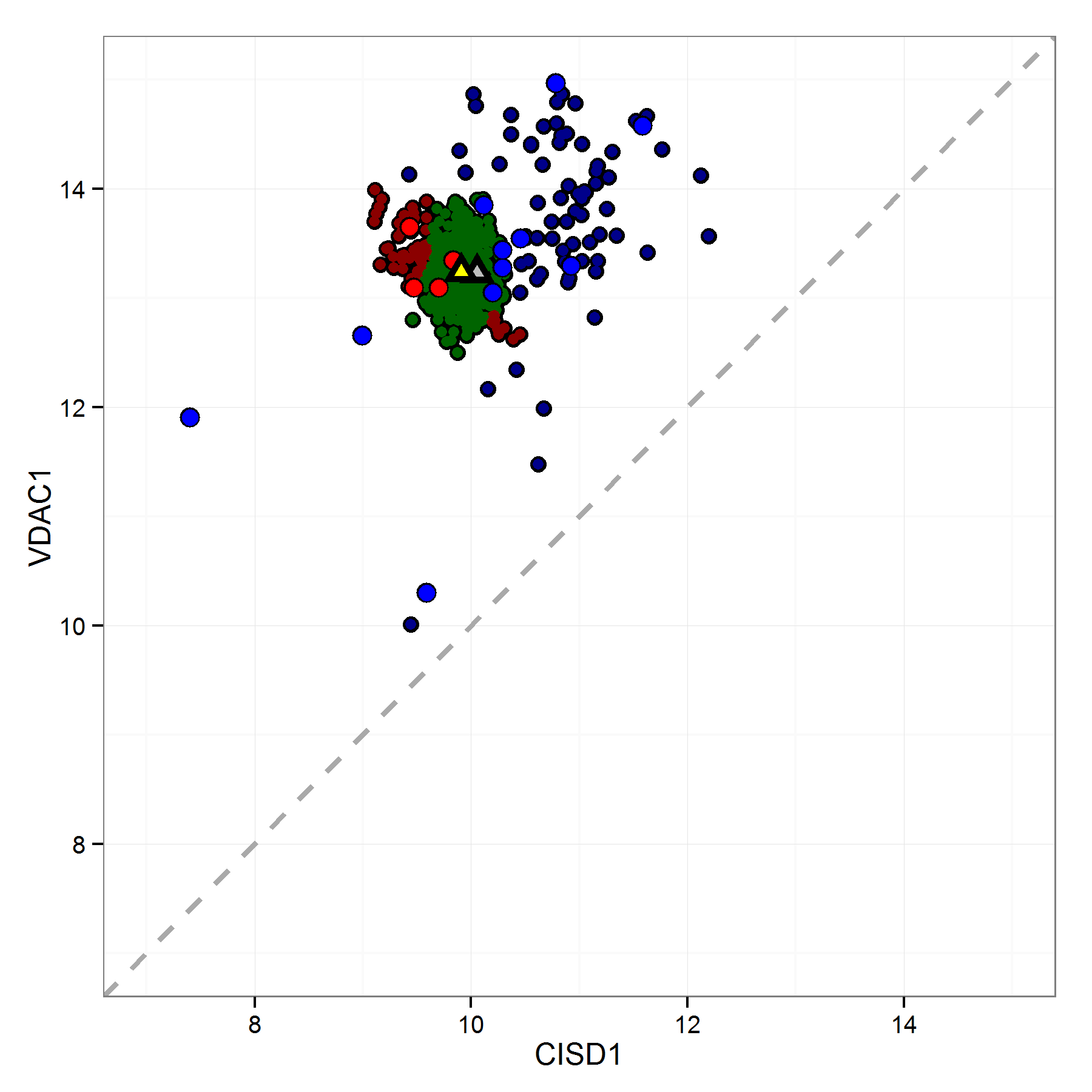}
  \caption[Results of MCDC on LINCS L1000 data - example 2]{Example 2
    showing results of applying MCDC to L1000 data. On the left is the
    data before correction, and on the right is that data after
    correction. Triangles indicate inferred mean - gray is the mean of
    all the data while yellow is the mean of the largest cluster found
    by MCDC.}
  \label{fig:example2}
\end{figure}

\begin{figure}[ht]
  \centering
  \includegraphics[width=.48\textwidth]{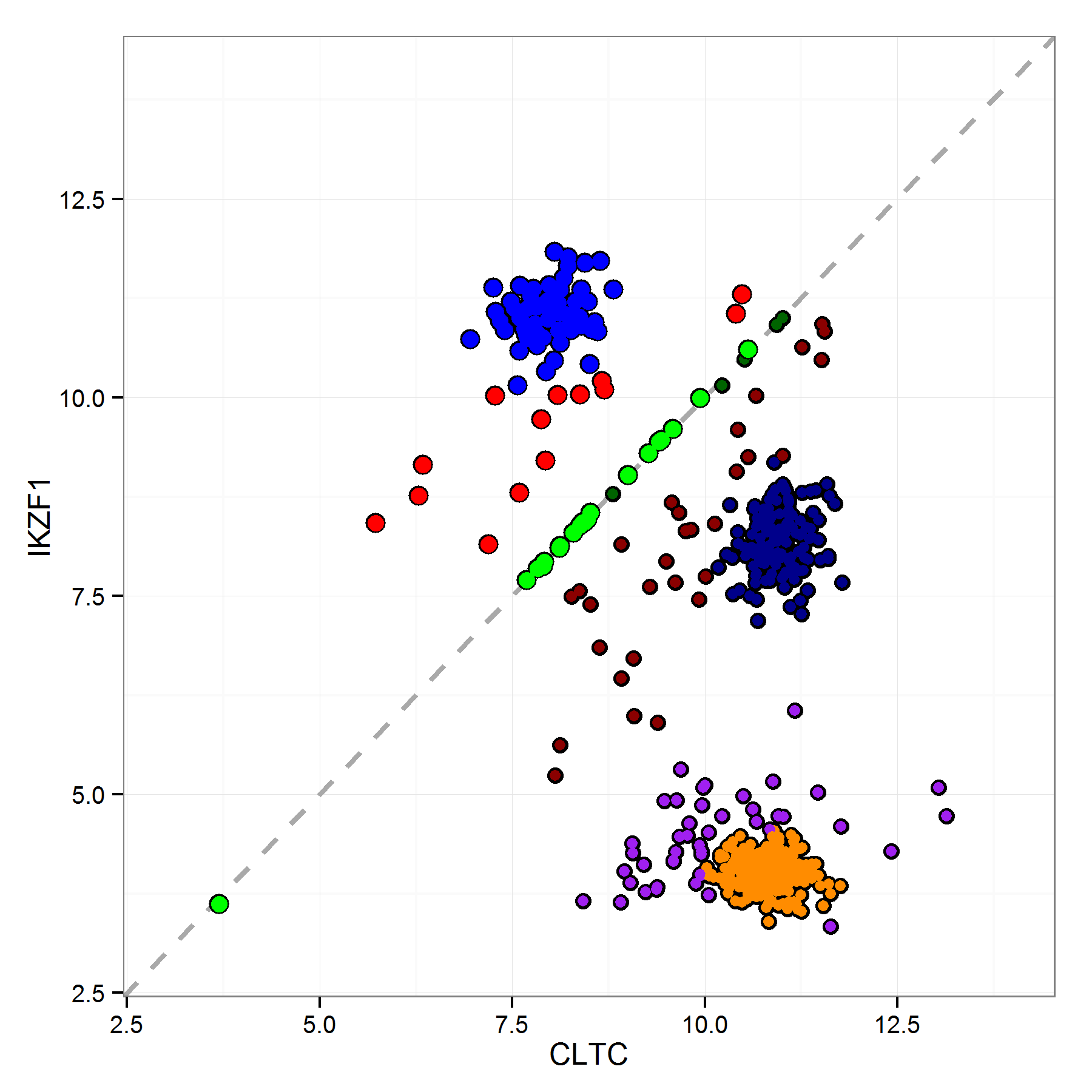}
  \includegraphics[width=.48\textwidth]{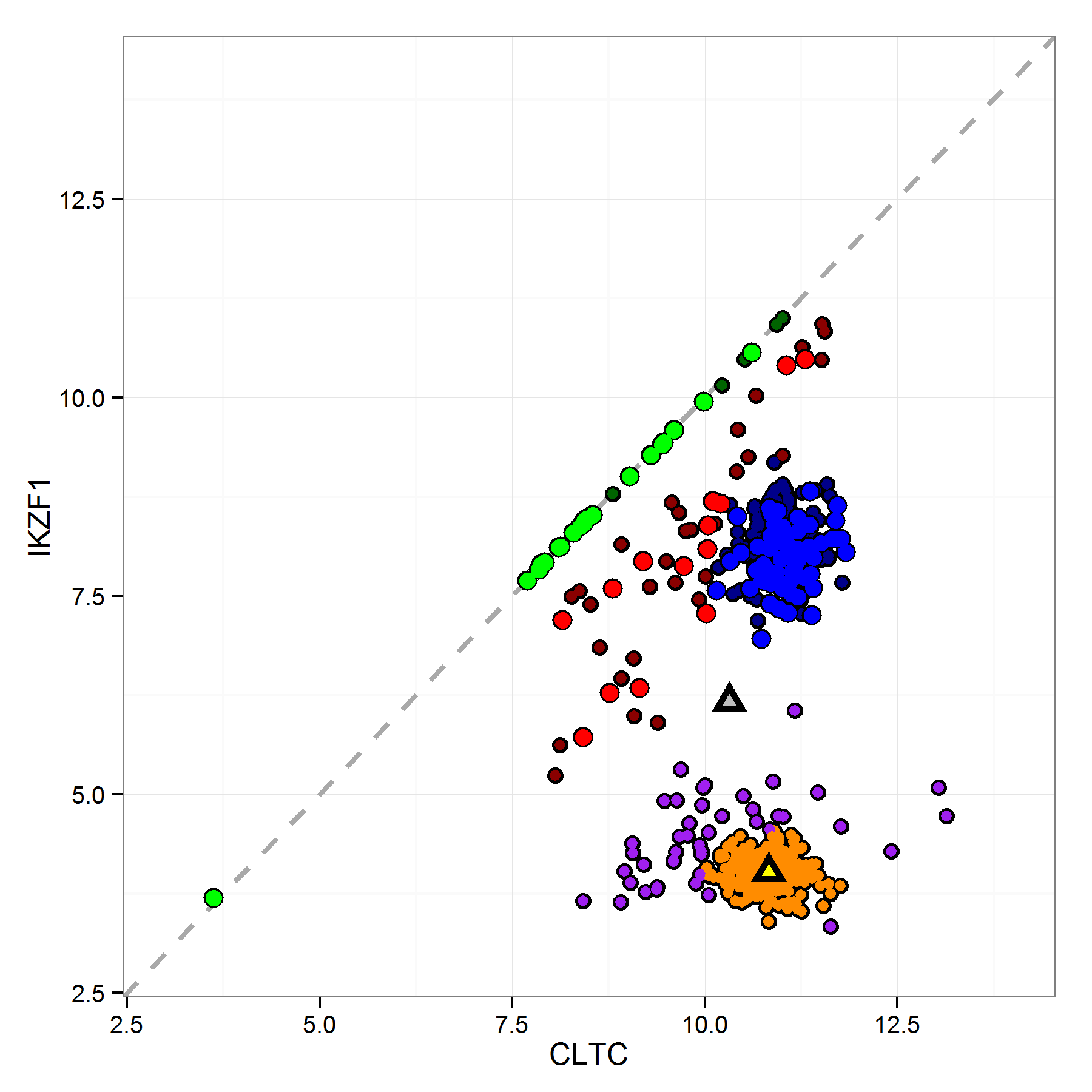}
  \caption[Results of MCDC on LINCS L1000 data - example 3]{Example 3
    showing results of applying MCDC to L1000 data. On the left is the
    data before correction, and on the right is that data after
    correction. Triangles indicate inferred mean - gray is the mean of
    all the data while yellow is the mean of the largest cluster found
    by MCDC.}
  \label{fig:example3}
\end{figure}

Figures \ref{fig:example2} and \ref{fig:example3}
show MCDC applied to additional gene pairs in the same dataset. In
each case MCDC succeeded in removing the artifacts in the data.

\subsection{Agreement with External Baseline Data}

First of all, we wanted to see if MCDC improves the data relative to
an external baseline. There are about 2,000 untreated experiments in
the A375 cell line.
These experiments should all yield similar expression
levels since they are all done under the same experimental
conditions. We can get an estimate of the gene expression level of a
particular gene by taking the mean across all the experiments. We
refer to this as the unaltered data.

There are two expression level baseline datasets included in the LINCS L1000
metadata for cell line A375, one using RNAseq technology and the
other using Affymetrix microarray technology. Each of these data were
generated using an independent technology and can be compared to the
values in the L1000 data. Since the baseline datasets are produced
using different technologies, the scales of the expression levels are
different than that from the L1000 data. In order to take this
difference into account, we looked at the mean squared error (MSE)
from a simple linear regression of the baseline data on the 
inferred gene expression levels from the L1000 data.

We then applied MCDC to see if this improved the estimates of gene
expression. To do this, we applied MCDC separately to each pair of
genes that were measured using the same bead color. For a gene pair,
we ran MCDC on the data from the 2,000 experiments. Doing this for
all 500 gene pairs, we ended up with an estimated gene expression
level for all of the 1,000 landmark genes. These estimates were also
compared to the baseline estimates and we were able to compare the
MSEs of the unaltered data with those from the corrected data.  Note
particularly in Figure \ref{fig:example3} that the inferred mean after
MCDC is substantially different than the mean of the full dataset,
moving it from a location not near any data to within the largest
cluster.

Table \ref{tab:mseComp} shows the results of this analysis. Using the
corrected data improved the MSE by 8\% when using the Affymetrix data
and by 7\% when using the RNAseq data. 

\begin{table}[ht]
  \caption[MSE results from MCDC applied to LINCS L1000 data]{MSE of
    regressing external baseline data on imputed gene
    means. Comparison of unaltered means and MCDC data. Affymetrix and 
    RNAseq baselines both are from external sources independent of the
    LINCS L1000 data.}
  \centering
        \begin{tabular}[ht]{l|cc}
          & \textbf{Affymetrix} & \textbf{RNAseq} \\
          \textbf{Method} & \textbf{Baseline} & \textbf{Baseline} \\
          \hline
          \textbf{Unaltered} & 1.91 & 1.66 \\
          \textbf{MCDC} & 1.76 & 1.55 \\ \hline
        \end{tabular}
  \label{tab:mseComp}
\end{table}

\subsection{Gene Regulatory Network Inference}

As well as improving the overall estimates of gene expression levels,
MCDC identifies particular experiments where the
gene pairs are flipped. This improvement of the data leads to
improvements in methods that use the data in a more granular way.

To see this improvement, we looked at inferring gene regulatory
networks from the LINCS L1000 data. We looked at knockdown
experiments, which target a specific gene to suppress its expression
level. This target gene is the regulator in these experiments, and the
remaining genes are potential targets, giving us a causal pathway by
which to infer networks.

We developed a simple posterior probability approach using the
knockdown data to infer edges. To do this, we first standardized the
knockdown data using the untreated experiments on the same plate to
get $z$-values. We then used a simple linear regression model, regressing each
potential target on the knocked down gene. This can be converted into
a posterior probability of there being an edge from the knocked-down
gene to the target gene. This approach is fast and allows us to
incorporate prior probabilities as well. The final result is a ranked
edgelist.

The LINCS L1000 data includes multiple knockdown experiments for most
of the landmark genes. We used the posterior probability method on the
knockdown experiments for cell line A375 to generate a ranked list of
potential edges. In order to assess the results, we used a gene-set
library compiled from TRANSFAC and JASPAR \citep{wingender2000,
  sandelin2004} and accessed from Enrichr at
\url{http://amp.pharm.mssm.edu/Enrichr/} \citep{chen2013enrichr}. This
is a list of transcription factors, namely genes that are known to
control the expression levels of other genes. Each transcription
factor has a list of target genes, yielding an assessment edgelist.

The TRANSFAC and JASPAR (T\&J) edgelist is not a complete reference
since not all gene relationships are captured in the T\&J
library. This is partially because the T\&J data focuses on
transcription factors but also because the true regulatory networks
are not fully known. The T\&J edgelist includes edges for 37
transcription factors also found among the LINCS landmark genes. This
includes approximately 4,200 regulator-target pairs out of about
42,000 potential edges for which we computed posterior probabilities.

To see the effect of MCDC, we applied the posterior probability method
using the unaltered data and compared the results with using the
same posterior probability method on the data after it had been corrected using MCDC. This
results in two ranked lists of gene pairs with associated posterior
probabilities. We compare these with the T\&J assessment dataset by
taking all edges with a posterior probability over a specified cutoff
and creating two-by-two tables showing how well the truncated
edgelists overlap with the T\&J edgelist

Table \ref{tab:2x2} shows the two-by-two tables generated at posterior
probability cutoffs of 0.5 and 0.95. We also report approximate
$p$-values by using the probability of getting at least the number of
true positives found using a binomial($n$, $p$) distribution, where
$n$ is the number of pairs in the inferred list and $p$ is the
probability of selecting a true edge from the total number of possible
edges. From the table, we can see that the $p$-value is better for the
corrected data at both probability cutoffs. The corrected data
includes more edges at both cutoffs but maintains a similar precision,
the proportion of edges which are true edges.

\begin{table}[ht]
  \centering
  \caption[2x2 tables for applying MCDC to knockdown experiments]{
    2x2 tables for cell line A375 using knockdown experiments
    for finding edges. Compared to TRANSFAC and JASPAR from Enrichr. When
    using the unaltered data and looking at edges with posterior
    probability of 0.5 or greater, 41 of the 302 candidate edges are found
    in TRANSFAC and JASPAR, and 14 of the 81 candidate edges at a cutoff
    of 0.95 are true edges. Similarly, when using the MCDC-corrected data,
    63 of the 463 candidate edges at a cutoff of 0.5 are true edges and 20
    of the 119 at a cutoff of 0.95 are true edges. Approximate binomial
    p-values included.
  }
  \begin{tabular}[ht]{lrlcclcc}
    & & & \multicolumn{5}{c}{\textbf{T\&J}} \\
    & & & \multicolumn{2}{c}{\textbf{cutoff: 0.5}} & &
    \multicolumn{2}{c}{\textbf{cutoff: 0.95}} \\
    & & & Yes & No & & Yes & No \\
    \cline{4-5} \cline{7-8}
    & \multirow{2}{*}{\textbf{Unaltered}}
    & Yes & \multicolumn{1}{|c|}{41} & \multicolumn{1}{c|}{261} & &
    \multicolumn{1}{|c|}{14} & \multicolumn{1}{c|}{67} \\
    \cline{4-5} \cline{7-8}
    & & No & \multicolumn{1}{|c|}{4152} & \multicolumn{1}{c|}{38836} & &
    \multicolumn{1}{|c|}{4179} & \multicolumn{1}{c|}{39030} \\
    \cline{4-5} \cline{7-8}
    & & & \multicolumn{2}{l}{p-value: 0.02} & &
    \multicolumn{2}{l}{p-value: 0.02} \\ 
    ~ \\
    \cline{4-5} \cline{7-8}
    &  \multirow{2}{*}{\textbf{MCDC}} & Yes & \multicolumn{1}{|c|}{63}
    & \multicolumn{1}{c|}{400} & & \multicolumn{1}{|c|}{20} &
    \multicolumn{1}{c|}{99} \\ 
    \cline{4-5} \cline{7-8}
    & & No & \multicolumn{1}{|c|}{4130} & \multicolumn{1}{c|}{38697} & &
    \multicolumn{1}{|c|}{4173} & \multicolumn{1}{c|}{38998} \\
    \cline{4-5} \cline{7-8}
    & & & \multicolumn{2}{l}{p-value: 0.004} & &
    \multicolumn{2}{l}{p-value: 0.01}
  \end{tabular}
  \label{tab:2x2}
\end{table}

Another way of looking at the results is via the precision-recall
curve \citep{Raghavan&1989}. Precision and recall are both calculated
by truncating our ranked list of edges and looking only at the edges
in the truncated list. Precision is the proportion of the edges in the
truncated list which are true edges. Recall is the proportion of all
true edges which are in the truncated list. The precision-recall curve
takes a ranked list of edges from a procedure and shows how the
precision varies as more and more edges are included from that
list. High precision at low recall indicates that the procedure is
good at identifying true edges at the highest probability. This is
important in many cases, particularly genetic studies, because it
gives researchers good information on where to focus their efforts in
subsequent studies.

\begin{figure}[ht]
  \centering
  \includegraphics[width=.6\textwidth]{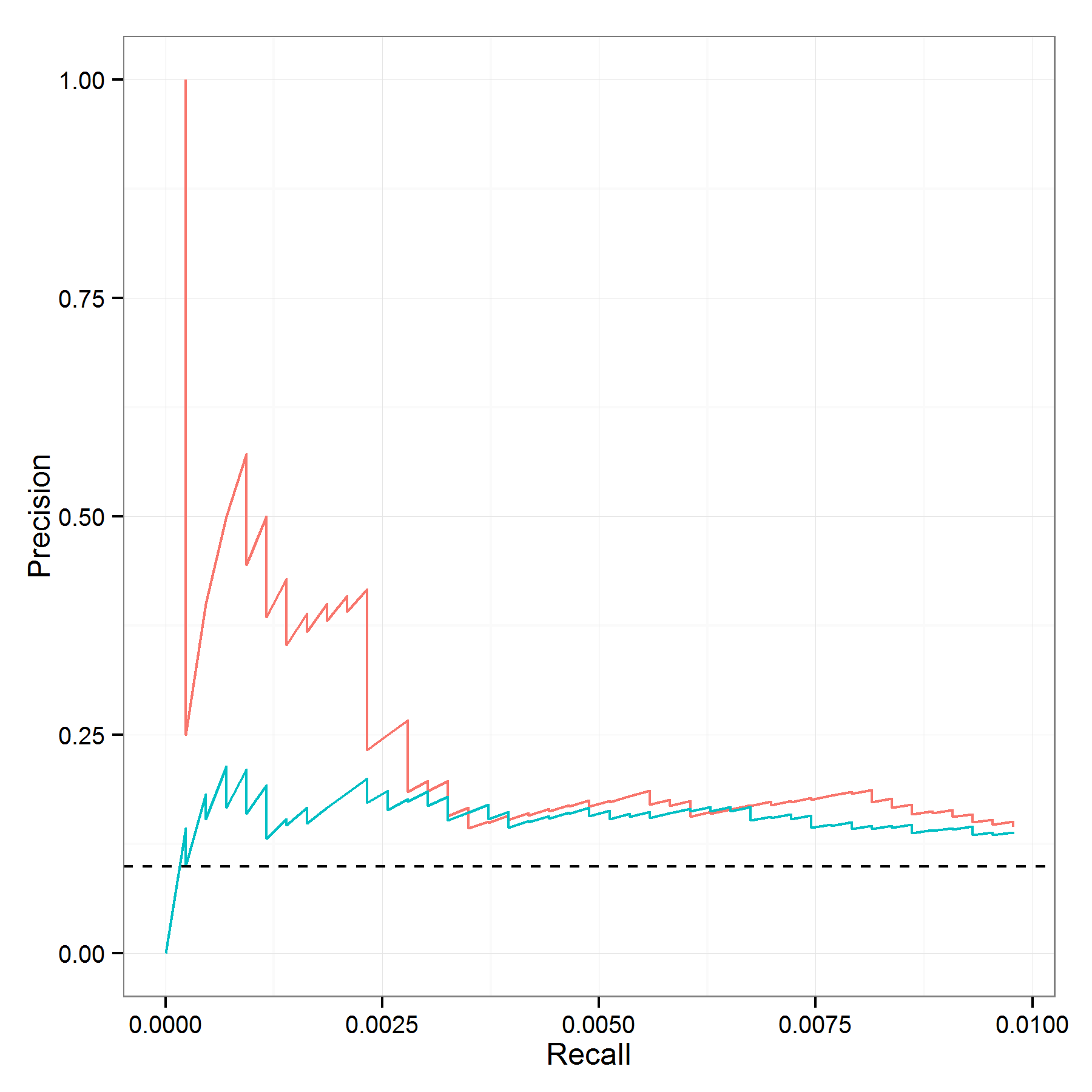}
  \caption[Precision-recall curve comparing unaltered and
  MCDC-corrected data]{Precision-recall curve comparing edgelists from
    unaltered (blue line) and MCDC-corrected (red line) data on knock
    down data.}
  \label{fig:prc}
\end{figure}

Figure \ref{fig:prc} compares the precision-recall curves for the
unaltered and corrected data at the very top of the edgelist. The
dashed line shows what would be expected by randomly ordering the
edges, and above that line is an improvement. Both methods are
improvements, but the corrected data yields much better results for
the very top edges returned. This is of particular importance for
further research because having high confidence in the top edges
allows the biologist to develop further experiments to focus on these
edges in additional, more targeted experiments. In this respect, data
correction with MCDC provides substantial improvement.

We can see this by looking at the inferred edges, ranked so that the
first edge has the highest posterior probability, the second has the
second highest, and so one. Table \ref{tab:topgenes} is constructed by
ranking the edgelist from the posterior probability method on a
particular dataset. Thus the first edge in the list is that with the
highest posterior probability, the second edge has the next highest
posterior probability, and so on. We then look at each edge and see if
it also is found in the T\&J assessment edgelist. The rank in the
table indicates the position at with the $n$-th edge in T\&J was found in
the ranked edgelist. So the edge with highest posterior probability
using the MCDC-corrected data is in T\&J, as is the edge with the 5th
highest posterior probability, etc. Only one of the top 10 edges from
the unaltered data is a true edge, while 5 of the top 10 edges from
the MCDC-corrected data are true edges.

\begin{table}[ht]
  \caption[Results of applying MCDC to knockdown data - top
  edges]{Comparison of the rank of the first 5 edges found that 
    match the TRANSFAC and JASPAR edgelist. Edges ranked by posterior
    probability. MCDC-corrected data produces found edges at higher
    ranks than the uncorrected data. See text for explanation of how
    the table was constructed.}
  \centering
  \begin{tabular}[ht]{c|cc}
    \textbf{Found Edge} & \textbf{Unaltered Rank} & \textbf{MCDC Rank} \\
    \hline
    \textbf{1} & 7 & 1 \\
      \textbf{2} & 11 & 5 \\
      \textbf{3} & 14 & 6 \\
      \textbf{4} & 19 & 7 \\
      \textbf{5} & 26 & 10 \\ \hline
  \end{tabular}
  \label{tab:topgenes}
\end{table}

\section{Discussion}  \label{s6}

When working with any data, understanding the unique aspects of how it
was generated and processed can be helpful in developing models and
methods, leading to improved inference. This is particularly true with
genomic data. There are often many steps of data transformation and
normalization between the raw measured data and what is used by the
researcher in drawing conclusions \citep{binder2008}. When these steps
are not known or understood, assumptions about sources of error can be
misinformed and lead to degraded performance in
inference. \cite{price2006} identified population stratification of
allele frequency in disease studies, while \cite{gomez2009} found that
a particular sequencing technique resulted in many artificial
replicates. \cite{lehmann2013} showed that quantile normalization of
microarray data introduced a phase shift into time-series in strains
of cyanobacteria, changing night-expressed genes into day-expressed
genes and vice versa. \cite{stokes2007} developed a tool to identify
and remove artifacts in genomic data. Batch effects have been
identified as a significant source of systematic error that can be
corrected \citep{leek2010,chen2011,sun2011}. Identifying these sources
of error is crucial, and in some cases can lead to vastly improved
results.

We showed how understanding the data-processing pipeline of the LINCS
L1000 data allowed us to identify the introduction of a particular error,
namely the flipping of expression values for gene pairs. This led to
the development of MCDC, which is able to identify and correct these
flipping errors. We were able to apply MCDC to improve the L1000 data
in aggregate, as measured against external standards. This improvement
of the data also led to improved inference of regulatory relationships
between the genes, in particular for the edges ranked
highest. Moreover, the use of the EM algorithm for optimization makes
MCDC fast and useful for large datasets.

MCDC is an extension of model-based clustering,
which has been extensively used in other analyses of genetic data, 
including sequence analysis
\citep{verbist2015} and image analysis of microarrays
\citep{li2005}. One of the most common uses of model-based clustering
in genetics is in finding meaningful groups among gene expression
profiles across multiple experiments under different experimental
conditions (cell sources, phases, applied drugs, etc.)
\citep{siegmund2004,jiang2004}. This includes methods using
Gaussian mixture models \citep{yeung2001}, infinite mixture models
\citep{medvedovic2002} and Bayesian hierarchical clustering
\citep{cooke2011}. Our use of MCDC as a step in improving data
quality is complementary to these analysis methods.

We showed in our simulation experiments that MCDC is able to
accurately identify the data points which have been altered and thus
improve the quality of the data. It is not limited to flipping as seen
in the LINCS data, but is able to handle any dataset where a subset of
the data points have been altered in a known way. One possible
extension of this method would be to compare different possible data
transformations to identify which is most compatible with the observed
data.

\bibliographystyle{apalike}
\bibliography{references}

\end{document}